\documentclass[12pt]{article}

\usepackage[margin=1in]{geometry}

\usepackage{amsthm,amsmath,amsfonts,amssymb}
\usepackage{graphicx,caption,subcaption}
\usepackage{enumitem}

\usepackage{multicol,longtable,multirow,booktabs}
\usepackage{color,hyperref,xcolor}
\hypersetup{colorlinks=true,urlcolor=blue,citecolor=blue}
\usepackage[authoryear]{natbib}
\usepackage{mathaddons} 

\title{\bf Robust Rank Estimation for Noisy Matrices}
\author{Subhrajyoty Roy
    \\
    Department of Statistics and Data Science\\
    Washington University at St. Louis, USA\\
    \and
    Abhik Ghosh \\
    Interdisciplinary Statistical Research Unit\\
    Indian Statistical Institute, Kolkata, India\\
    \and 
    Ayanendranath Basu \\
    Interdisciplinary Statistical Research Unit\\
    Indian Statistical Institute, Kolkata, India
}
\date{}

\begin{document}
\maketitle 

\bigskip
\begin{abstract}
  Estimating the true rank of a noisy data matrix is a fundamental problem underlying techniques such as principal component analysis, matrix completion, etc. Existing rank estimation criteria, including information-based and cross-validation methods, are either highly sensitive to outliers or computationally demanding when combined with robust estimators. This paper proposes a new criterion, the Divergence Information Criterion for Matrix Rank (DICMR), that achieves both robustness and computational simplicity. Derived from the density power divergence framework, DICMR inherits the robustness properties while being computationally very simple. We provide asymptotic bounds on its overestimation and underestimation probabilities, and demonstrate first-order B-robustness of the criteria. Extensive simulations show that DICMR delivers accuracy comparable to the robustified cross-validation methods, but with far lower computational cost. We also showcase a real-data application to microarray imputation to further demonstrate its practical utility, outperforming several state-of-the-art algorithms.
\end{abstract}

\noindent%
{\it Keywords:} Robust Principal Component Analysis, Model Selection, Minimum Divergence Estimation.

\section{Introduction}\label{sec:rank-intro}

Numerous statistical tools have been developed to analyze the structure of a dataset represented in a matrix form. A common challenge in applying some of these tools, such as the Singular Value Decomposition (SVD), Principal Component Analysis (PCA), Non-negative Matrix Factorization (NMF), Matrix Completion (MC), etc. is the determination of the true rank of the data matrix. In some applications, this true rank can be guessed based on the domain knowledge or external data (e.g. rank is usually taken as 1 or 2 in the problem of video background modelling~\citep{roy2024robustsvd}), while some studies completely avoid them~\citep{choi2025inference}. However, in most applications, this rank is not known and must be estimated in a data-dependent fashion~\citep{greenacre2022principal}. For example, in various genome-wide studies, one often works with a microarray data which is partially observed~\citep{moorthy2019survey}. To impute these missing values, a standard technique is to assume a low-rank structure of the entire data matrix, estimate this low-rank component based on observed entries and use that to derive the imputed values. In this case, the quality of imputation is greatly affected by the choice of the rank of this low-rank component. Although the number of patients or clusters of cells may be known from exogeneous datasets, the relationships between various genes are not known during collection of the microarray data (e.g., RNA sequencing), which limits the ability to guess the true rank beforehand. Later in Section~\ref{sec:rank-real-application}, we demonstrate one such example of how the method proposed in this paper can be used to obtain a robust estimate of the rank of the low-rank component to solve the microarray data imputation problem.

There are primarily three branches of methods for the determination of the rank: subjective and ad hoc methods (e.g. scree plot), distribution based methods (e.g. information criterion, and thresholding) and computational methods (e.g. cross-validation); see~\cite{jolliffe_principal_2002} for a summary. Unfortunately, none of these existing methods are resistant to outliers. As will be shown later in Section~\ref{sec:rank-sim}, a naive approach of replacing the classical SVD or PCA procedures with robust SVD or PCA procedures either does not provide a reliable estimate of the rank, or significantly increases the computational cost. In this paper, we aim to address this gap by proposing a novel rank estimation method which is both robust and computationally simple.

\subsection{Mathematical Description of the Problem}

Before we present the existing methods for determination of the matrix rank, let us first present the mathematical formulation of the problem. We are provided with an $n \times p$ data matrix $\bb{X}$ that decomposes into a low-rank component $\bb{L}$, a sparse component $\bb{S}$ and a dense perturbation component $\bb{N}$ as 
\begin{equation}
    \bb{X} = \bb{L} + \bb{S} + \dfrac{\sigma}{\sqrt{ \max\{n, p\} }} \bb{N}.
    \label{eqn:rank-lsn-decomp}
\end{equation}
\noindent Here, the noise matrix $\bb{N}$ is composed of i.i.d. entries with some unknown density function $g$ such that $\E(n_{ij}) = 0$ and $\var(n_{ij}) = 1$ for all $i = 1, 2, \dots, n$ and $j = 1, 2, \dots, p$ and $\sigma$ is the common noise variance. Given the data matrix $\bb{X}$, the objective is then to robustly estimate $r = \rank{\bb{L}}$. The decomposition given in~\eqref{eqn:rank-lsn-decomp} arises in many applications, such as robust PCA~\citep{roy2024robustpca}, robust SVD~\citep{roy2024robustsvd}, NMF~\citep{chicoki2011robustnmf}, clustering analysis~\citep{du2017robust}, image processing~\citep{sadek2012svd}, etc. 

It is worthwhile to note that the scaling factor of $1/\sqrt{\max\{n,p\}}$ in the noise component is necessary to ensure identifiability of this rank estimation problem. As shown in~\cite{candes2011robust}, if the noise variance does not asymptotically go to $0$ as either $n$ or $p$ increases to infinity, the largest singular value of the noise matrix $\bb{N}$ and its corresponding singular vectors can be added to $\bb{L}$ to constitute a new low-rank component with a different rank than that of $\bb{L}$. Additionally, \cite{shabalin2013reconstruction} proved that any reasonable matrix decomposition technique cannot consistently estimate the singular values of $\bb{L}$ (or its functions like the rank, i.e., the number of nonzero singular values) from the data matrix $\bb{X}$ if the noise variance is larger than the order of $1/\sqrt{np}$. The same scaling factor also appears in Assumption (B6) employed in the proof of the consistency of the rSVDdpd estimator~\citep{roy2024robustsvd}, which we shall make use of later.

\subsection{Organization of the paper}

In this paper, we propose a novel matrix rank estimation criterion called divergence information criterion for matrix rank (DICMR) based on the divergence information criterion (DIC) by~\cite{karagrigoriou2008measures}. The criterion is derived in Section~\ref{sec:rank-dicmr}. Prior to that, we briefly review various existing rank estimation techniques in Section~\ref{sec:rank-literature}. Section~\ref{sec:rank-theory} deals with establishing theoretical properties of the proposed criterion, including selection consistency and influence function analysis. The proposed method is resistant to outlying observations, achieves a performance similar to (if not better than) various cross-validation methods, while remaining computationally simple and efficient. This has been demonstrated through extensive simulation studies in Section~\ref{sec:rank-sim}. Finally, in Section~\ref{sec:rank-real-application}, we showcase a real use-case of the proposed method for the microarray data imputation problem and also compare it against several existing imputation algorithms. The proposed method is found to be superior in accuracy than these other existing methods. Due to the space constraints, all proofs of the theoretical results are provided in Section~\ref{appendix:rank-proof} of the Appendix.

\section{Existing Literature}\label{sec:rank-literature}

\subsection{Penalization Methods}\label{sec:rank-penalized}

As discussed in Section~\ref{sec:rank-intro}, one of the major classes of techniques used to determine the rank of a matrix is distribution-based methods such as information criteria, thresholding, etc. These methods are usually built by combining a statistical discrepancy measure of fit from a partial SVD estimate to the data matrix $\bb{X}$ and a penalty function of the rank used in the partial SVD estimate. Hence, we refer to such methods as penalization methods. As one includes higher rank components in the estimate, the error in estimation reduces by yielding a small discrepancy measure, but the penalty function compensates by being an increasing function of the rank. Mathematically, this approach considers a model selection criterion of the form
\begin{equation}
    C(r) = d(\bb{X} - \bbhat{L}^{(r)}) + g(r), \ r = 0, 1, 2, \dots, \min(n,p),
    \label{eqn:rank-penalized-C-function}
\end{equation}
\noindent where $d(\bb{A})$ is an appropriate measure of the norm of the matrix $\bb{A}$ and $\bbhat{L}^{(r)}$ is the estimate of the low-rank component $\bb{L}$ based on a partial SVD estimate up to rank $r$. The quantity $g(r)$ is an appropriately chosen penalty function of the rank $r$. The estimate of rank is then given by 
\begin{equation*}
    \widehat{r} = \argmin_{0 \leq r \leq \min\{n, p\}} C(r).   
\end{equation*}
\noindent Usually, $d(\cdot)$ is taken to the squared Frobenius norm scaled appropriately by the estimated noise variance $\widehat{\sigma}^{(r)}$ (i.e., the estimated variance of the entries of the error matrix $\bb{E}$) and $\bbhat{L}^{(r)}$ is given by $\sum_{k=1}^r \widehat{\lambda}_k\widehat{u}_{ki}\widehat{v}_{kj}$, where $\widehat{\lambda}_k, \bbhat{u}_k$ and $\bbhat{v}_k$ are the estimates of the $k$-th largest singular value and corresponding left and right singular vectors. Classical model selection criteria like Akaike's Information Criterion (AIC) \citep{akaike1973information} and Bayesian Information Criterion (BIC)~\citep{schwarz1978estimating} are usually inadequate for this purpose, since the penalty functions $g(r)$ in this case follows an inverted parabola form like $r(n+p-r)$, leading to overestimation. Thus, \cite{bai2002determining} proposed six modified criteria named PC1, PC2, PC3, IC1, IC2 and IC3, all of which use different linearly increasing functions of $r$ as the penalty functions.

Thresholding methods estimate the rank as the maximum index for which the estimated singular value (or principal component eigenvalue) exceeds a certain threshold. This threshold is often determined based on various distributional assumptions of the error matrix $\bb{E}$. Although the process looks different, it can be shown to be a special case of the penalization approach by considering the criterion function
\begin{equation*}
    C(r) = \sum_{k > r} \widehat{\lambda}_k + r \Vert \bb{X}\Vert_F \ind{\{\lambda_r \leq \tau\} },
\end{equation*}
\noindent where $\tau$ is the specific threshold, $\Vert\bb{X}\Vert_F$ denotes the Frobenius norm and $\ind{A}$ denotes the indicator function of the event $A$. Some recent prominent thresholding based methods are by~\cite{shabalin2013reconstruction,choi2017selecting} and~\cite{xu2021adaptive}. Among these, only~\cite{xu2021adaptive} incorporate a robustness component by using a robust PCA method in their algorithm, but without any provable statistical or robustness guarantees.

\subsection{Cross-Validation Approaches}\label{sec:rank-cv}

Another major approach in estimating the rank of a matrix uses a resampling technique known as cross-validation. In this approach, the rows and the columns of the data matrix $\bb{X}$ are subdivided into groups; these groups are deleted in respective turns. The remaining entries are then used to compute a partial SVD estimate up to a chosen rank, and it is used to predict the deleted entries. The choice of the rank that yields the lowest possible prediction error is declared to be the estimate of the rank. Mathematically, let $R_1, R_2, \dots, R_B$ be $B$ randomly chosen subsets of the rows having indices from $\{1, 2, \dots, n\}$ and similarly, let $C_1, C_2, \dots, C_B$ be such subsets of the columns having indices from $\{1, 2, \dots, p\}$. The choice of $B$ may depend on the dimensions of the matrix. Typically, $B = \binom{n}{n_1} \binom{p}{p_1}$ for some $n_1 \leq n$ and $p_1 \leq p$. For any choice of $b = 1, 2, \dots, B$, the data matrix $\bb{X}$ is then partitioned into
\begin{equation*}
    \bb{X} = \begin{bmatrix}
        \bb{X}_{R_b, C_b} & \bb{X}_{R_b, C_b^c}\\
        \bb{X}_{R_b^c, C_b} & \bb{X}_{R_b^c, C_b^c}\\
    \end{bmatrix}.
\end{equation*}
\noindent The tuple $(R_b, C_b)$ is often called the holdout set following the terminology in the cross-validation literature. Based on this partitioned matrix, one then finds an estimate of $\bb{X}_{R_b, C_b}$ using partial SVD up to rank $r$ of the other entries, 
\begin{equation}
    \bbhat{X}^{(r)}_{R_b, C_b} = T_r\left( \bb{X}_{R_b, C_b^c}, \bb{X}_{R_b^c, C_b},  \bb{X}_{R_b^c, C_b^c} \right),
    \label{eqn:rank-T-function}
\end{equation}
\noindent where $T_r$ denotes the prediction function based on the partial SVD estimate. Finally, we combine the error in estimation to create a cross-validation criterion
\begin{equation}
    \text{CV}(r) := S\left( \left\{ \norm{\bb{X}_{R_b, C_b} - \bbhat{X}_{R_b, C_b}^{(r)} } \right\}_{b = 1}^B \right),    \ r = 1, 2, \dots, \min\{n, p\},
    \label{eqn:rank-cv-function}
\end{equation}
\noindent where $S(\cdot)$ is a suitable measure of the scale of univariate samples. If the rank $r$ is the true rank, then the distribution of these errors should be homogeneous across all such random choice of partitions, hence yielding a small value of the cross-validation metric given in~\eqref{eqn:rank-cv-function}. Based on this idea, the estimated rank is obtained as 
\begin{equation*}
    \widehat{r} = \displaystyle\argmin_{0 \leq r \leq \min\{n,p\} } \text{CV}(r).
\end{equation*}

\cite{wold1978crossvalidation} considers ``speckled''-type holdout sets where the sets $R_b$ and $C_b$ are singleton sets covering each row and column indices, resulting in $B = np$ holdout sets. The prediction function $T_r$ based on the partial SVD estimates is calculated in an iterative fashion based on Expectation-Maximization (EM) algorithm. For a fixed choice of holdout set combination $(R_b, C_b)$, it starts with an imputation of the entry $X_{R_b, C_b}$ as the mean of remaining entries and performs a partial $r$-SVD of the entire imputed matrix. The entry $X_{R_b, C_b}$ is then imputed by its revised estimate from this partial $r$-SVD and then a further partial SVD is performed on this new imputed matrix. These two steps are repeated until for two successive runs the imputed value of $X_{R_b, C_b}$ remains stable. Finally, the scale measure for calculating the cross-validation error is simply given by the root mean squared error (RMSE).

On the other hand, \cite{gabriel2002lebiplot} considers a ``block''-type holdout set where the sets $R_b$ and $C_b$ are all possible fixed-size sets from all possible combinations of row and column indices. His proposed estimate $T_r$ is given by
\begin{equation*}
    T_r\left( \bb{X}_{R_b, C_b^c}, \bb{X}_{R_b^c, C_b},  \bb{X}_{R_b^c, C_b^c} \right)
    =  \bb{X}_{R_b, C_b^c} \left( \bb{X}_{R_b^c, C_b^c} \right)^{(r), +} \bb{X}_{R_b^c, C_b},
\end{equation*}
\noindent where $\left( \bb{X}_{R_b^c, C_b^c} \right)^{(r), +}$ is a generalized inverse of the partial rank $r$-SVD of $\bb{X}_{R_b^c, C_b^c}$, given by 
\begin{equation*}
    \left( \bb{X}_{R_b^c, C_b^c} \right)^{(r), +} = \sum_{k=1}^r \dfrac{1}{\widehat{\lambda}_{k, 0}} \bbhat{u}_{k, 0} \bbhat{v}_{k, 0}\tr,
\end{equation*}
\noindent where $\widehat{\lambda}_{k, 0}, \bbhat{u}_{k, 0}, \bbhat{v}_{k, 0}$ are classical SVD estimates of singular values and vectors based on the partitioned matrix $\bb{X}_{R_b^c, C_b^c}$ only. While the initial proposal by~\cite{gabriel2002lebiplot} considers only singleton holdout sets, \cite{owen2009bi} demonstrated that generalizing the holdout sets to $n_r \times n_c$-size partitions (where $n_r, n_c > 1$) leads to a more reliable estimate of the rank for large data matrices. Their recommendation was to use $n_r = [n/2]$ and $n_c = [p/2]$ to have the best bias-variance trade-off. However, this increases the number of holdout sets from $np$ to $\mathcal{O}(n^{n_r}p^{n_c})$, significantly increasing the time complexity of the entire cross-validation procedure.

In another direction, \cite{eastment1982crossvalidation} consider the holdout for rows and columns separately. For a fixed $b$, they choose $R_b = C_{b+1} = \phi$, the null set and $R_{b+1} = \{ i_0 \}$ and $C_b = \{ j_0 \}$, where $i_0$ and $j_0$ are randomly chosen row and column indices. Subsequently, let us denote $\widehat{\lambda}_{k, 0}^{(b)}, \bbhat{u}_{k, 0}^{(b)}, \bbhat{v}_{k, 0}^{(b)}$ as the SVD estimates of the $n \times (p-1)$ data matrix after deleting $j_0$-th column and  $\widehat{\lambda}_{k, 0}^{(b+1)}, \bbhat{u}_{k, 0}^{(b+1)}, \bbhat{v}_{k, 0}^{(b+1)}$ as the same for $(n-1) \times p$ data matrix after deleting $i_0$-th row. They are then combined to produce an estimate of $X_{i_0, j_0}$ as 
\begin{equation}
    \widehat{X}_{i_0, j_0}^{(r)} = \sum_{k=1}^r \left( \widehat{\lambda}_{k, 0}^{(b)} \right)^{1/2} \left( \widehat{\lambda}_{k, 0}^{(b+1)} \right)^{1/2} \bbhat{u}_{k, 0}^{(b)} \left( \bbhat{v}_{k, 0}^{(b+1)} \right)\tr.
    \label{eqn:rank-cv-rowcol-1}
\end{equation}
\noindent Note that, since both of these estimates ignore the entry $X_{i_0, j_0}$ either by removing row $i_0$ or column $j_0$, Eq.~\eqref{eqn:rank-cv-rowcol-1} yields a proper cross-validation estimate of $X_{i_0, j_0}$ based on partial SVD up to rank $r$. However, \cite{bro2008cv} performed extensive simulation studies and found that it is useful to add a scaling factor to the singular value estimates for small matrices, such as
\begin{equation}
    \widehat{X}_{i_0, j_0}^{(r)} = \sum_{k=1}^r \left( \widehat{\lambda}_{k, 0}^{(b)} \sqrt{p/(p-1)}  \right)^{1/2} \left( \widehat{\lambda}_{k, 0}^{(b+1)}  \sqrt{n/(n-1)} \right)^{1/2} \bbhat{u}_{k, 0}^{(b)} \left( \bbhat{v}_{k, 0}^{(b+1)} \right)\tr.
    \label{eqn:rank-cv-rowcol-2}
\end{equation}
\noindent Asymptotically, this scaling factor goes to $1$, hence both variants of cross-validation algorithms of~\cite{eastment1982crossvalidation} yield similar performances for large matrices.

\subsection{Other Approaches}\label{sec:rank-other}

Beyond these two main bodies of works, other ad hoc approaches have also been proposed in the literature. The simplest one is the ``Elbow'' method, conceptualized by~\cite{thorndike1953elbow}, which proceeds by looking at the estimated singular values (or the squared estimated eigenvalues) of the data matrix $\bb{X}$ and finding the position in the screeplot where a major slope change occurs in the graph of singular values, leading to the shape of an elbow. In a different direction, \cite{hoff2007model} proposed a matrix rank estimation method using Bayesian modelling of the data matrix $\bb{X}$. His approach considers von-Mises distribution for the singular vectors and uses the properties of this distribution to derive the posterior distribution of the rank. Gibb's sampling is then used to generate the posterior samples. 

\section{Divergence Information Criterion for Matrix Rank}\label{sec:rank-dicmr}

\cite{roy2024robustsvd} recently proposed a robust SVD estimation procedure called ``rSVDdpd'' for estimating the low rank component $\bb{L}$ in the decomposition~\eqref{eqn:rank-lsn-decomp}. It aims to minimize the density power divergence (DPD)~\citep{basu1998robust} between the empirical density estimate and a model density $f$ governing the behaviour of the error component $\bb{E} := \bb{S} + \sigma \bb{N} / \sqrt{\max(n,p)}$. The corresponding objective function is given by
\begin{multline}
    H_{\alpha}^{(r)}(\bb{\theta}; \bb{X}) = \dfrac{1}{np}\sum_{i=1}^n \sum_{j=1}^p V_{ij,\alpha}(\bb{\theta})\\
    = \dfrac{1}{np}\sum_{i=1}^n \sum_{j=1}^p  \sigma^{-\alpha}\left[ \int f^{1+\alpha} - \left( 1 + \dfrac{1}{\alpha} \right) f^\alpha\left( \left\vert  \dfrac{X_{ij} - \sum_{k=1}^r \lambda_k u_{ki}v_{kj} }{\sigma} \right\vert \right) \right].
    \label{eqn:rank-dpd-objective}
\end{multline}
\noindent Here, the parameter $\bb{\theta}$ consists of the singular values $\lambda_k$ and corresponding left and right singular vectors $\bb{u}_k := (u_{k1}, \dots, u_{kn})\tr$ and $\bb{v}_k := (v_{k1}, \dots, v_{kp})\tr$ for $k = 1, 2, \dots, r$, and the noise variance $\sigma^2$. The parameter $\alpha$ is a robustness parameter used to balance the efficiency and the robustness of the above estimator, and is usually restricted in the unit interval $[0, 1]$. As $\alpha \rightarrow 0$, the objective~\eqref{eqn:rank-dpd-objective} transforms into an equivalent formulation based on the Kullback-Leibler (KL) divergence; minimization of which recovers the classical maximum likelihood estimate of SVD. Thus, the rSVDdpd estimator is a robust generalization of the classical SVD estimator, controlled by the robustness parameter $\alpha$. Let us denote $\bbhat{\theta}_\alpha^{(r)} = ( \bbhat{D}_\alpha^{(r)}, \bbhat{U}_\alpha^{(r)}, \bbhat{V}_\alpha^{(r)}, (\widehat{\sigma}_\alpha^{(r)})^2 )$ as the rSVDdpd estimator of the parameter $\bb{\theta}$ when the rank is restricted to $r$ and the robustness parameter is $\alpha$. To obtain $\bbhat{\theta}_\alpha^{(r)}$ by minimizing of the objective function given in~\eqref{eqn:rank-dpd-objective}, \cite{roy2024robustsvd} noticed that if the left (right) singular vectors $\bbhat{U}_\alpha^{(r)}$ ($\bbhat{V}_\alpha^{(r)}$) are known, then the estimation of rest of the parameters is equivalent to solving a linear regression problem. Based on this observation, they proposed an iterative algorithm based on alternating regression approach to obtain the rSVDdpd estimate, and proved the convergence of the algorithm and statistical consistency of the converged estimator under some regularity conditions.

For the limiting case $\alpha \to 0$, if the errors are modelled by Gaussian densities with mean zero, then
\begin{equation}
    H_0^{(r), \phi}(\bb{\theta}) := \lim_{\alpha \rightarrow 0} H_\alpha^{(r),\phi}(\bb{\theta}) = (2\sigma^2)^{-1} \dfrac{1}{np}\sum_{i=1}^n \sum_{j=1}^p \left( X_{ij} - \sum_{k=1}^r \lambda_k u_{ki} v_{kj} \right)^2,
    \label{eqn:rank-H-function-zero}
\end{equation}
\noindent where the superscript $\phi$ indicates that $H(\cdot)$ is calculated based on the standard normal density. Eq.~\eqref{eqn:rank-H-function-zero} establishes that the limiting form of the discrepancy measure $H(\cdot)$ relates to the usual Frobenius norm used in the classical criteria like AIC, BIC, etc. Based on this observation, \cite{karagrigoriou2008measures} proposed a ``Divergence Information Criterion'' (DIC) based on the density power divergence~\citep{basu1998robust} which is demonstrated to be robust in selecting the correct model in the presence of outliers. For the matrix rank estimation problem, the DIC is given as
\begin{equation}
    \text{DIC}_\alpha(r) = H_\alpha^{(r)}(\bbhat{\theta}_\alpha^{(r)}) + r(\alpha+1)(2\pi)^{-\alpha/2}\left( \dfrac{1+\alpha}{1+2\alpha} \right)^{3/2},
    \label{eqn:rank-DIC}
\end{equation}
\noindent where $\bbhat{\theta}_\alpha^{(r)}$ denotes the ``rSVDdpd'' estimator up to rank $r$ for the robustness tuning parameter $\alpha$.

A more general form of the DIC for the independent but non-identically distributed setup is developed by~\cite{kurata2018robust}, who proposed a new criterion BHHJ-C given as
\begin{equation}
    \text{BHHJ-C}_\alpha = H_\alpha^{(r)}(\bbhat{\theta}_\alpha^{(r)}) + \dfrac{1}{n}\trace{J_\alpha^{-1}(\bbhat{\theta}_\alpha^{(r)}) K_\alpha(\bbhat{\theta}_\alpha^{(r)})},
    \label{eqn:rank-bhhj-c}
\end{equation}
\noindent where  
\begin{align*}
    J_\alpha(\bb{\theta}) & = \dfrac{1}{np}\sum_{i=1}^n \sum_{j=1}^p \E_{g_{ij}}\left[ \nabla^2_{\bb{\theta}} V_{ij,\alpha}^{(r)}(\bb{\theta})  \right],\\
    K_\alpha(\bb{\theta}) & = \dfrac{1}{np}\sum_{i=1}^n \sum_{j=1}^p \E_{g_{ij}}\left[ \left( \nabla_{\bb{\theta}} V_{ij,\alpha}^{(r)}(\bb{\theta} ) \right) \left( \nabla_{\bb{\theta}} V_{ij,\alpha}^{(r)}(\bb{\theta}) \right)\tr \right],
\end{align*}
\noindent where $V_{ij,\alpha}(\cdot)$ is the $V$-function defined in~\eqref{eqn:rank-dpd-objective}, $g_{ij}$ is the true density of the entries $X_{ij}$ of the matrix, and $\nabla_{\bb{\theta}}$ and $\nabla^2_{\bb{\theta}}$ respectively denote the gradient and the Hessian operator with respect to the parameter $\bb{\theta}$. Since $g_{ij}$s are unknown, one computes the above expectations assuming that the densities belong to the model families of densities, and then replacing $\bb{\theta}$ by its consistent estimate $\bbhat{\theta}_\alpha^{(r)}$ to compute the BHHJ-C.

We begin with a derivation similar to the derivation of DIC as in~\cite{karagrigoriou2008measures}. The objective is to find a robust unbiased estimator of $\E(Q_{\alpha}^{(r)}(\bb{\theta}) \mid \bb{\theta} = \bbhat{\theta}_\alpha^{(r)} )$, where
\begin{equation}
    Q_{\alpha}^{(r)}(\bb{\theta}) := \dfrac{1}{np}\sum_{i=1}^n \sum_{j=1}^p \int V_{ij,\alpha}^{(r)}(\bb{\theta}; x) g_{ij}(x)dx,
    \label{eqn:rank-Q-function}
\end{equation}
\noindent and $V_{ij,\alpha}^{(r)}(\cdot)$ are as defined in~\eqref{eqn:rank-dpd-objective} and $g_{ij}$s are the true densities of the entries $X_{ij}$ of the data matrix. Note that, the $Q_{\alpha}^{(r)}(\cdot)$ function is simply the population version of the DPD objective function $H_\alpha^{(r)}(\cdot)$ given in~\eqref{eqn:rank-dpd-objective}. As for the rSVDdpd estimator, conditional on the event $\bb{V}_\alpha^{(r)} = \bbhat{V}_\alpha^{(r)}$, the estimation of the other parameters of $\bbhat{\theta}_\alpha^{(r)}$ is same as solving a robust linear regression problem, we can use the BHHJ-C criterion as in~\eqref{eqn:rank-bhhj-c} to obtain $\E(Q_{\alpha}^{(r)}(\bb{\theta}) \mid \bb{V}_\alpha^{(r)} = \bbhat{V}_\alpha^{(r)} )$. Proceeding with a derivation similar to that of~\cite{ghosh2013robust}, we can calculate the $J_\alpha(\bb{\theta})$ and $K_\alpha(\bb{\theta})$ matrices for a general model family $f$. This yields the penalty term of the BHHJ-C conditioned on $\bb{V}_\alpha^{(r)} = \bbhat{V}_\alpha^{(r)}$ as
\begin{equation}
    \trace{J_\alpha^{-1}(\bbhat{\theta}_\alpha^{(r)}) K_\alpha(\bbhat{\theta}_\alpha^{(r)})}
    = \trace{ \dfrac{ \left(\widehat{\sigma}_\alpha^{(r)}\right)^{-(2\alpha+2)} C^f_{2\alpha} }{ \left(\widehat{\sigma}_\alpha^{(r)}\right)^{-(\alpha+2)} C^f_{\alpha} } \bb{I}_r }\\
    = \trace{ \left(\widehat{\sigma}_\alpha^{(r)}\right)^{-\alpha} C^f_{2\alpha}/C^f_\alpha  },
    \label{eqn:rank-jk-penalty}
\end{equation}
\noindent where $\widehat{\sigma}^2$ is the estimated noise variance and
\begin{equation}
    C^f_\alpha = \int (f'(\vert x\vert))^2 f^{\alpha - 1}(\vert x\vert) dx,
    \label{eqn:rank-Cf-defn}
\end{equation}
\noindent assuming $f'$ exists. In the special case of the normal model family, we have $f(x) = \phi(x) = (2\pi)^{-1/2}e^{-x^2/2}$ and hence
\begin{equation*}
    C^\phi_\alpha = (2\pi)^{-\alpha/2} (1+\alpha)^{-3/2}.
\end{equation*}
\noindent Note that, here we also use the orthogonality of the estimated right singular vectors $\bbhat{V}_\alpha^{(r)}$ to conclude that $(\bbhat{V}_\alpha^{(r)})\tr \bbhat{V}_\alpha^{(r)} = \bb{I}_r$, the identity matrix, and as a result, the penalty term in~\eqref{eqn:rank-jk-penalty} turns out to be free of $\bbhat{V}_\alpha^{(r)}$. Using this approach, we obtain
\begin{align*}
    \E\left( Q_{\alpha}^{(r)}(\bbhat{\theta}_\alpha^{(r)}) \right)
    & = \dfrac{1}{p}\sum_{j=1}^p \E\left[ \dfrac{1}{n}\sum_{i=1}^n \E_{g_{ij}} V_{ij}^{(r)}(\bbhat{\theta}_{\alpha}^{(r)}; X_{ij}) \right]\\
    & = \dfrac{1}{p}\sum_{j=1}^p \E\left[ \E\left( \dfrac{1}{n}\sum_{i=1}^n \E_{g_{ij}} V_{ij}^{(r)}(\bbhat{\theta}_{\alpha}^{(r)}; X_{ij}) \right) \mid \bb{V}_\alpha^{(r)} = \bbhat{V}_\alpha^{(r)} \right]\\
    & = \dfrac{1}{p}\sum_{j=1}^p \E\left[ \dfrac{1}{n}\sum_{i=1}^n V_{ij}^{(r)}(\bbhat{\theta}_\alpha^{(r)}; X_{ij}) + \dfrac{r}{n} \left(\widehat{\sigma}_\alpha^{(r)}\right)^{-\alpha} C^f_{2\alpha}/C^f_{\alpha} + o(1/n) \right], \ \text{ by~\eqref{eqn:rank-jk-penalty},}\\
    & = \E\left[ H_{\alpha}^{(r)}(\bbhat{\theta}_\alpha^{(r)}, \bb{X}) + \dfrac{r}{n} \left(\widehat{\sigma}_\alpha^{(r)}\right)^{-\alpha} C^f_{2\alpha}/C^f_{\alpha} + o(1/n) \right].
\end{align*}
\noindent Here, $o(1/n)$ refers to a term that tends to $0$ faster than $1/n$ as $n \to \infty$. We can similarly take conditional expectation conditioned on the left singular vectors to obtain
\begin{equation*}
    \E\left( Q_{\alpha}^{(r)}(\bbhat{\theta}_\alpha^{(r)}) \right) = \E\left[ H_{\alpha}^{(r)}(\bbhat{\theta}_\alpha^{(r)}, \bb{X}) + \dfrac{r}{p} \left(\widehat{\sigma}_\alpha^{(r)}\right)^{-\alpha} C^f_{2\alpha}/C^f_{\alpha} + o(1/p) \right].
\end{equation*}
\noindent Now since both $n$ and $p$ tend towards infinity, and if $r = \Ocal(\min(n, p))$, we obtain an estimate by averaging the above two unbiased estimators, resulting in a new criterion given as
\begin{equation}
    \text{DICMR}_\alpha(r) := H_{\alpha}^{(r)}(\bbhat{\theta}_\alpha^{(r)}, \bb{X}) + \dfrac{r (n+p)}{2np} \left(\widehat{\sigma}_\alpha^{(r)}\right)^{-\alpha} C^f_{2\alpha}/C^f_{\alpha}.
    \label{eqn:rank-dicmr}
\end{equation}
\noindent We shall refer to this as the ``Divergence Information Criterion for Matrix Rank'' (DICMR) in the subsequent discussion. In the special case of normally distributed errors, the DICMR reduces to
\begin{equation}
    \text{DICMR}^\phi_\alpha(r) := H_{\alpha}^{(r),\phi}(\bbhat{\theta}_\alpha^{(r)}, \bb{X}) + \dfrac{r (n+p)}{2np} (2\pi)^{-\alpha/2} \left(\widehat{\sigma}_\alpha^{(r)}\right)^{-\alpha} \left( \dfrac{1+\alpha}{1+2\alpha} \right)^{3/2}.
    \label{eqn:rank-dicmr-normal}
\end{equation}

At this point, it may be useful to compare the form of the DICMR with various existing criteria. Table~\ref{tab:rank-penalized-methods} provides such a summary of various penalized criteria adapted to the matrix rank estimation.

\begin{table}[htbp]
    \centering
    \caption{Different Penalized Criteria for Matrix Rank Estimation. (Here $\widehat{\sigma}_\alpha$ denotes the estimated noise variance by minimizing the objective function $H_\alpha(\cdot)$ as in~\eqref{eqn:rank-dpd-objective}).}
    \label{tab:rank-penalized-methods}
    \resizebox{\textwidth}{!}{
    \begin{tabular}{ll}
        \toprule
        \textbf{Criterion} & \textbf{Definition}\\
        \midrule
        AIC~\citep{akaike1973information} &  $2 H_0^{(r),\phi}(\bbhat{\theta}_0^{(r),\phi})\widehat{\sigma}_0^2 + r \widehat{\sigma}_0^2 (n + p - r)/np$ \\
        BIC~\citep{schwarz1978estimating} & $2 H_0^{(r), \phi}(\bbhat{\theta}_0^{(r),\phi})\widehat{\sigma}_0^2 + r \widehat{\sigma}_0^2 (n + p - r) \log(np)/np$ \\
        PC1~\citep{bai2002determining} & $2 H_0^{(r), \phi}(\bbhat{\theta}_0^{(r), \phi})\widehat{\sigma}_0^2 + r \widehat{\sigma}_0^2 \log(\frac{n+p}{np})/np$\\
        PC2~\citep{bai2002determining} & $2H_0^{(r), \phi}(\bbhat{\theta}_0^{(r),\phi})\widehat{\sigma}_0^2 + r \widehat{\sigma}_0^2 \log(\min(n, p))/np$\\
        PC3~\citep{bai2002determining} & $2H_0^{(r),\phi}(\bbhat{\theta}_0^{(r), \phi})\widehat{\sigma}_0^2 + r \widehat{\sigma}_0^2 \log(\min(n, p))/\min(n,p)$\\        
        IC1~\citep{bai2002determining} & $2H_0^{(r),\phi}(\bbhat{\theta}_0^{(r), \phi})  + r \log(\frac{n+p}{np})/np$\\
        IC2~\citep{bai2002determining} & $2H_0^{(r), \phi}(\bbhat{\theta}_0^{(r),\phi}) + r  \log(\min(n,p))/np$\\
        IC3~\citep{bai2002determining} & $2H_0^{(r), \phi}(\bbhat{\theta}_0^{(r), \phi})  + r \log(\min(n,p))/\min(n, p)$\\
        DIC~\citep{karagrigoriou2008measures} & $H_\alpha^{(r)}(\bbhat{\theta}_\alpha^{(r)}) + r (\alpha + 1)(2\pi)^{-\alpha/2}\left( \frac{1+\alpha}{1+2\alpha} \right)^{1+r/2}$\\
        RCC~\citep{kurata2020consistency} & $H_\alpha^{(r)}(\bbhat{\theta}_\alpha^{(r)}) + r \log(n)/2n$ \\
        DICMR (ours) & $H_\alpha^{(r)}(\bbhat{\theta}_\alpha^{(r)}) + \frac{r (n+p)}{2np} (2\pi)^{-\alpha/2}\widehat{\sigma}_\alpha^{-\alpha} \left( \frac{1+\alpha}{1+2\alpha} \right)^{3/2}$\\
        \bottomrule
    \end{tabular}}
\end{table}

\section{Theoretical Studies}\label{sec:rank-theory}

\subsection{Selection Consistency}\label{sec:rank-selection-consistency}

A key property of a model selection criterion is its selection consistency. It refers to the property that if two competing models are present; one of which is nested under the other, the criterion selects the smaller model if it is adequate (i.e., contains the true distribution) and chooses the larger model if the smaller model is not adequate, asymptotically with probability tending to $1$. In the case of the rank estimation problem, the selection consistency for the DICMR implies that if the low rank component $\bb{L}$ of the data matrix $\bb{X}$ as in~\eqref{eqn:rank-lsn-decomp} is of rank $r$, then the corresponding criterion should asymptotically satisfy 
\begin{equation*}
    \prob\left( \text{DICMR}_\alpha(r) \leq \min\{ \text{DICMR}_\alpha(r-1), \text{DICMR}_\alpha(r+1) \} \right) \rightarrow 1,
\end{equation*}
\noindent for any $r \geq 1$ as both $n$ and $p$ tend to infinity, subject to the restriction that $\lim_{n,p\rightarrow \infty}(n/p) = c$ for some $c \in (0, \infty)$. In the following discussions, we will aim to investigate this property for the DICMR criterion defined in~\eqref{eqn:rank-dicmr}. However, for simplicity, we shall restrict our attention to two completing models of rank zero and rank one. As the DICMR criterion uses rSVDdpd estimator~\cite{roy2024robustsvd} as its backbone, which proceeds by sequentially estimating the singular values and vectors one by one, consideration of this special case barely limits the general applicability of the results presented in this section. These analogus results will continue to hold for other choices of rank $r \geq 1$, provided that the key assumptions are modified appropriately. 

We will begin by considering the situation when the low rank component $\bb{L}$ is of rank one, i.e., the entries $X_{ij} = \lambda_1 u_{1i}v_{1j} + \sigma_1 e_{ij}$ for all $i = 1, 2,\dots, n$ and $j = 1, 2, \dots, p$ where $e_{ij}$s are independent and identically distributed according to the model density $f$ and the variance $\sigma_1^2 = 1/\sqrt{np}$. In this case, a consistent criterion would select the larger model of rank one since the smaller model with rank zero is not adequate. The following theorem establishes this fact for the DICMR.

\begin{theorem}\label{thm:rank-selection-consistency-one}
    Let us assume that the Assumptions (B1)-(B6) of~\cite{roy2024robustsvd} hold with the density $g$ replaced by the model density $f$. Additionally, assume that
    \begin{enumerate}
        \item The model density $f$ is bounded.
        \item The estimated noise variance $(\widehat{\sigma}_\alpha^{(0)})^2$ under rank zero assumption satisfies $\sigma_1 / \widehat{\sigma}_\alpha^{(0)} \xrightarrow{P} 0$ asymptotically as both $n$ and $p$ tend to infinity, where $\sigma_1^2$ is the true noise variance.
    \end{enumerate}
    \noindent Then, for any $\alpha > 0$, we have
    \begin{equation*}
        \lim_{\substack{n \rightarrow \infty, p\rightarrow \infty\\(n/p) \rightarrow c \in (0, \infty)}} P\left( \textnormal{DICMR}_\alpha(0) > \textnormal{DICMR}_\alpha(1) \right) = 1.
    \end{equation*}
\end{theorem}

It is useful to note that the assumption on the ratio of the noise variance to be asymptotically negligible stems from the assumption that the low rank component $\bb{L}$ of the data matrix $\bb{X}$ is of rank one. However, as explained earlier, it is impossible to determine this rank unless the singular values are significantly larger than a threshold, depending on $n, p$ and the noise variance $\sigma_1^2$. Thus, we require the estimated noise variance $(\widehat{\sigma}_\alpha^{(0)})^2$ under rank zero assumption to be significantly larger than the true noise variance $\sigma_1^2$, since the former will combine effects from both the noise variance and the first singular value. Also, the use of robust rSVDdpd estimate $(\widehat{\sigma}_\alpha^{(0)})^2$ here avoids the non-identifiability issue pointed out by~\cite{candes2011robust}. For instance, the matrix $\bb{L} = \bb{e}_1\bb{e}_1\tr$, where $\bb{e}_1$ is the first vector of the usual set of Euclidean basis vectors, is both rank $1$ and sparse, and hence, can be regarded as either a part of the low-rank component or the noise component. Using the robust rSVDdpd estimate $(\widehat{\sigma}_\alpha^{(0)})^2$ would ignore the effect of these entries as outlying observations and classify the above as part of the noise component.

Next, we will consider the situation when the low rank component $\bb{L}$ is of rank zero. This means the data matrix $\bb{X}$ consists of only pure noise such that they are independently distributed with zero mean and variance $\sigma_0^2 = 1/\sqrt{np}$. Then, the following theorem presents the probability that the DICMR criterion selects the rank zero model over the rank one model.

\begin{theorem}\label{thm:rank-selection-consistency-zero}
    Let us assume that the Assumptions (B1)-(B6) of~\cite{roy2024robustsvd} hold with the true density $g$ replaced by the model density $f$. Additionally, assume that 
    \begin{enumerate}
        \item The model density $f$ is bounded.
        \item In Assumption (B1), $\lambda^g = 0$, i.e., the low rank component of the data matrix $\bb{X}$ is $\bb{0}$.
    \end{enumerate}
    \noindent Then, for any fixed $\alpha > 0$, we have
    \begin{equation*}
        \lim_{\substack{n \rightarrow \infty, p\rightarrow \infty\\(n/p) \rightarrow c \in (0, \infty)}} \prob\left( \textnormal{DICMR}_\alpha(0) < \textnormal{DICMR}_\alpha(1) \right) \geq \dfrac{1}{2} + \dfrac{1}{2} \max\left\{ 0, 1 - t_\alpha^{-2} \right\}
    \end{equation*}
    \noindent where
    \begin{equation}
        t_\alpha = \dfrac{(1+c)}{2\sqrt{c}} \dfrac{C_{2\alpha}^f}{C_\alpha^f \norm{f}_{1+\alpha}^{1+\alpha} } \dfrac{B_\alpha^f}{\sqrt{B_{2\alpha}^f - (A_\alpha^f)^2} },
        \label{eqn:rank-t-func}
    \end{equation}
    \noindent and
    \begin{align}
        A_\alpha^f & = \int \left[ f(\abs{x}) + \abs{x} f'(\abs{x}) \right]f^\alpha(\abs{x})dx, \label{eqn:rank-A-alpha-func}\\
        B_\alpha^f & = \int f^{1+\alpha}(\abs{x})dx -2 \int \abs{x}f'(\abs{x})f^\alpha(\abs{x})dx + \int x^2 \left( f'(\abs{x}) \right)^2 f^{\alpha-1}(\abs{x})dx. \label{eqn:rank-B-alpha-func}
    \end{align}
\end{theorem}

Theorem~\ref{thm:rank-selection-consistency-zero} establishes that the DICMR criterion has a non-trivial probability of overestimation, i.e., selecting the rank one model over the rank zero model, even when the true low rank component $\bb{L}$ is of rank zero. This is also known in the case of linear regression models; see~\cite{kurata2018robust}. A careful analysis of the term $t_\alpha$ in~\eqref{eqn:rank-t-func} reveals that as $c$, the ratio of the dimensions, remain away from $1$, the probability of selecting the correct model, i.e., the rank zero model, increases. Therefore, the DICMR criterion is more reliable in selecting the correct model when the data matrix $\bb{X}$ is rectangular, compared to square matrices.

\begin{figure}[htbp]
    \centering
    \includegraphics[width=0.75\textwidth]{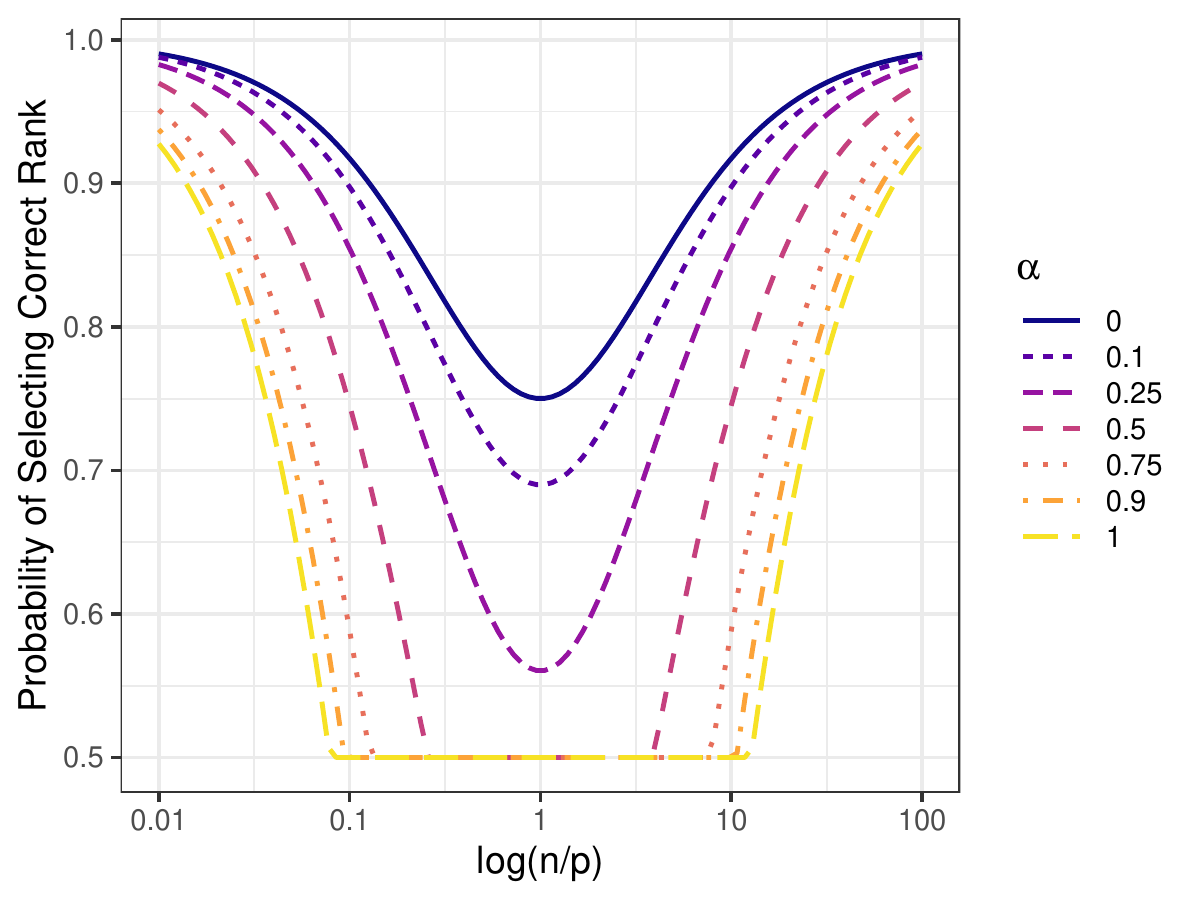}
    \caption{Probability of selecting the correct model (rank zero) as a function of the log-ratio $\log(n/p)$ for different values of $\alpha$ in the case of normally distributed errors.}
    \label{fig:correct-rank-prob-normal}
\end{figure}

In the case of normally distributed errors, the model density $f$ is the standard normal density $\phi(x) = (2\pi)^{-1/2}e^{-x^2/2}$ which is uniformly bounded by $(2\pi)^{-1/2}$. It is also easy to see that the Assumptions (B2)-(B3) of~\cite{roy2024robustsvd} hold for this case. As a result, under a reduced set of assumptions, both Theorem~\ref{thm:rank-selection-consistency-one} and Theorem~\ref{thm:rank-selection-consistency-zero} hold for the DICMR criterion defined in~\eqref{eqn:rank-dicmr-normal}. For this special case, the quantity $t_\alpha$ is given by
\begin{equation*}
    t_\alpha = \dfrac{(1+c)}{2\sqrt{c}} \dfrac{(2+\alpha^2)}{(1+\alpha)^{7/2}} \left[ \dfrac{(2+4\alpha^2)}{(1+2\alpha)^{5/2}} - \dfrac{\alpha^2}{(1+\alpha)^3} \right]^{-1/2}.
\end{equation*}
\noindent In Figure~\ref{fig:correct-rank-prob-normal}, we plot the lower bound to the probability of selecting the correct model as assured by Theorem~\ref{thm:rank-selection-consistency-zero} for the special case of normally distributed errors. As evident from the plot, as the robustness parameter $\alpha$ increases, one needs the ratio $c$ to be further away from $1$ to ensure that the DICMR criterion does not overestimate the rank of $\bb{L}$.

\subsection{Robustness}\label{sec:rank-robustness}

A popular measure of robustness for estimators is the influence function that measures the impact on an estimate due to the change in the distribution of the sample observations. In the case of a model selection criterion, various metrics have been proposed as a definition of influence function; see~\cite{toma2020robustmodel, machado1993robust}. However, we consider the definition as proposed by~\cite{kurata2022robustness} for our purpose. Let, $G_{ij}$ be the true distributions of the entries $X_{ij}$ of the data matrix, and let $K_{ij}$ be a collection of contaminating distributions for $i = 1, \dots, n; j = 1, \dots, p$. Define the $\epsilon$-contaminated distributions as
\begin{equation*}
    G_{\epsilon, ij} = (1-\epsilon)G_{ij} + \epsilon K_{ij}; \ i = 1, 2, \dots, n; j = 1, 2, \dots, p.
\end{equation*}
\noindent Under this setup, \cite{kurata2022robustness} proposes to measure the influence function of a model selection criterion $T(\cdot)$ as the limit 
\begin{equation}
    \text{IF}\left( T, \{ G_{ij} \}_{i,j=1}^{n,p} \right)
    = \lim_{\epsilon \to 0} \dfrac{1}{\epsilon} \left\vert T(\{ G_{\epsilon, ij} \}_{i,j=1}^{n,p}) -  T(\{ G_{ij} \}_{i,j=1}^{n,p}) \right\vert.
    \label{eqn:rank-inf-func}
\end{equation}
\noindent The model selection criterion is then first-order B-robust if the above limit in~\eqref{eqn:rank-inf-func} is finite. Before formally stating the result on robustness of DICMR, we make the following assumption.

\begin{assumption}\label{assum:integral-finite}
    The supremum of the expectations $\sup_{i,j} \E(f^\alpha((X_{ij}-\mu)/\sigma))$ and $\sup_{i,j} \E(\psi((X_{ij} - \mu)/\sigma))$ exist and are finite for any fixed $\mu$ and $\sigma$. Here, the expectations are taken with respect to $X_{ij} \sim G_{ij}$, $f$ is the model density, and $\psi(x) := f^{\alpha-1}(x) f'(x)/\vert x\vert$. 
\end{assumption}

Assumption~\ref{assum:integral-finite} is a very weak assumption which is satisfied in many cases. For example, if the model density $f$ and the true density $g_{ij}$s are uniformly $L^{1+\alpha}$-integrable, then an application of H\"{o}lder's inequality establishes the boundedness of the expectations given above. The following theorem now establishes that the proposed criterion DICMR is B-robust as per the definition of~\cite{kurata2020consistency}. 

\begin{theorem}\label{thm:dicmr-B-robust}
    Suppose that the model family of densities $f$ satisfies Assumptions (A1)-(A3) of~\cite{roy2024robustsvd} and additionally Assumption~\ref{assum:integral-finite} holds. Then, the model selection criterion DICMR as given in~\eqref{eqn:rank-dicmr} is first-order B-robust for any $\alpha > 0$, provided that the corresponding rSVDdpd estimator is B-robust.
\end{theorem}

Another aspect of measuring robustness is through breakdown point (BP) analysis. However, for the problem of rank estimation, such BP analysis is not appropriate because of the discrete nature of the parameter space. In this case, the parameter space of interest is given by $\Theta = \{ 0, 1, 2, \dots, \min(n,p) \}$ where $n$ and $p$ are the dimensions of the matrix, and its boundary is $\partial \Theta = \{ 0, \min(n,p) \}$. Now, consider a matrix of rank $2$. With modifications of a single entry of this matrix, one can only hope to modify its rank to either $1$ or $3$, but not to some value lying in the boundary $\partial\Theta$. Therefore, even the nonrobust rank estimators based on traditional SVD (traditional SVD has BP equal to $0$ for the singular value estimation problem) will not have a breakdown point equal to $0$. This renders the breakdown analysis uninformative for this particular estimation problem.

\section{Simulation Studies}\label{sec:rank-sim}

\subsection{Simulation Setups}\label{sec:rank-sim-setup}

To empirically verify the performance of the proposed DICMR, we will consider a simulation setup similar to the one described by~\cite{owen2009bi}. In each simulation setup, the data matrix is generated following the LSN decomposition as in~\eqref{eqn:rank-lsn-decomp}. To generate the low-rank component $\bb{L}$, we start by simulating an $n \times p$-dimensional matrix $\bb{A}$ with each entry $a_{ij}$ generated as an i.i.d. standard normal random variable. Then, we construct $\bb{L}$ as $\bb{L} = \bb{U}_{\bb{A}} \bb{D} \bb{V}_{\bb{A}}\tr$,  where $\bb{U}_{\bb{A}}, \bb{V}_{\bb{A}}$ are respectively the left and right singular vectors for the random matrix $\bb{A}$. The singular values of $\bb{L}$, i.e., the diagonal elements of $\bb{D}$ matrix are chosen in one of the two following ways: (a) \textbf{Equal Singular Values:} First $r$ singular values are taken to be equal to $1$ and rest of the singular values are kept $0$. (b) \textbf{Decreasing Singular Values:} The $r$ nonzero singular values are chosen to decrease linearly from $2$ to $1$. The entries $n_{ij}$ of the noise component $\bb{N}$ are independently generated according to a standard normal random variable. The noise variance $\sigma^2$ is chosen such that the noise-to-signal ratio is $\sigma_e^2$, i.e., $\E(\norm{\bb{N}}_F^2) = \sigma_e^2 \norm{\bb{L}}_F^2$. Finally, the entries $s_{ij}$ of the sparse component $\bb{S}$ are generated as 
\begin{equation*}
    s_{ij} = 5\delta_{ij}(2\eta_{ij}-1) \max_{i,j}\abs{L_{ij}}, \ i = 1, 2, \dots, n; j = 1, 2, \dots, p.
\end{equation*}
\noindent where $\delta_{ij}$ and $\eta_{ij}$ are independent Bernoulli random variables with success probabilities $\delta$ and $1/2$ respectively.

In the experiments, we consider $n = 50, p = 40$ and $r = 10$. We take different variants of the above setup with multiple combinations of noise-to-signal ratio $\sigma_e^2$ as $0.05, 0.5$ and $1$, and varying levels of contamination proportion $\delta$ ranging from $0$ (no contamination) to $0.05, 0.1$ and $0.2$ (high level of contamination). These simulation setups are indicated as $Sij$ with $i$-th level of contamination proportion and $j$-th level of noise-to-signal ratio. For example, (S12) indicate the simulation with noise-to-signal ratio $0.5$ and $5\%$ proportion of contamination. 

\subsection{Comparative Algorithms}

To measure comparative performances, we consider different rank estimation methods in addition to the proposed DICMR. Since the classical SVD estimates are highly influenced by the presence of outlying observations, most of the existing algorithms in literature are not resistant to outliers. The straightforward way to make these robust would be to swap the traditional SVD (or PCA) procedure used in these algorithms with a corresponding robust SVD procedure such as ``rSVDdpd''. However, even when we use robust SVD estimates to obtain $\bbhat{L}^{(r)}$ in~\eqref{eqn:rank-penalized-C-function}, the resulting rank estimator produces poor estimates due to using the squared Frobenius norm as the discrepancy measure $d(\cdot)$. The general divergence measures~\citep{karagrigoriou2008measures, toma2020robustmodel,kurata2022robustness}, and the proposed DICMR are more suited here.

Similarly, the cross-validation methods also produce poor estimates of the rank of a data matrix with outlying entries, when combined with traditional SVD. This is illustrated in detail through the simulation studies in Section~\ref{sec:rank-sim-result}. As in the case of penalized methods, we may replace the traditional SVD with the ``rSVDdpd'' procedure. But, it introduces significant computational bottlenecks as one needs to compute these partial robust SVD estimates for a large number of holdout sets. For instance, Gabriel style cross-validation of a $50 \times 40$-dimensional matrix requires computing $2000$ SVD operations. Even though ``rSVDdpd'' is quite fast compared to its peers as demonstrated by~\cite{roy2024robustsvd}, performing the robust SVD of a $50 \times 40$ matrix in a standard consumer-grade computer takes about a second, resulting in the rank estimation procedure taking approximately $33$ minutes in its entirety. This severely limits the practical usage of such a procedure. \cite{alarcon2022cv} considered a workaround where a robust SVD up to rank $\min\{n, p\}$ is used on the data matrix $\bb{X}$ first to estimate a full-rank proxy component $\bb{X}'$ free of outlying observations. Then, a cross-validation procedure based on the traditional SVD is applied on the proxy $\bb{X}'$, which is expected to recover the true rank. In our simulation studies, we consider this workaround for robust variants of cross-validation approaches.

\subsection{Simulation Results}\label{sec:rank-sim-result}

Due to space constraints, detailed results on various simulation setups are provided in the Appendix; only a summary is illustrated through Table~\ref{tab:rank-result-summary}. In particular, we demonstrate the best performing cross-validation method and the best performing penalized approach for each simulation scenario considered above.

\begin{table}[htbp]
    \centering
    \caption{Summarized results of the best performing rank estimation methods under different simulation scenarios (Column Prop. indicates the proportion of replications with exact estimation of correct rank; the method ``SVD'' refers to the traditional nonrobust SVD).}
    \label{tab:rank-result-summary}
    \resizebox*{\textwidth}{!}{
        \begin{tabular}{ccrrrrrrr}
            \toprule
            $\delta$ & $\sigma_e^2$ & \textbf{Singular} & \multicolumn{3}{c}{\textbf{Best CV Method}} & \multicolumn{3}{c}{\textbf{Best Penalty Method}}\\
             & \textbf{(1/SNR)} & \textbf{Values} & \textbf{Name} & \textbf{Prop.} & \textbf{RMSE} & \textbf{Name} & \textbf{Prop.} & \textbf{RMSE}\\
            \midrule
            \multirow{3}{*}{$0$} & $0.05$ & Equal & SVD + WCV, BCV & $1.00$ & $0.00$ & SVD + PC3, IC3 & $1.00$ & $0.00$   \\
             & $0.5$ & Equal & SVD + WCV, BCV & $1.00$ & $0.00$ & rSVDdpd + DICMR & $0.53$  &  $1.33$  \\
             & $1$ & Equal & rSVDdpd + WCV & $0.85$ & $0.39$ & rSVDdpd + DICMR & $0.39$ &  $1.28$  \\  
             \midrule
            \multirow{3}{*}{$0$} & $0.05$ & Decreasing & SVD + WCV & $1.00$ & $0.00$ & SVD + PC3, IC3 & $1.00$ &  $0.00$  \\
             & $0.5$ & Decreasing & SVD + ECV & $0.93$ & $0.32$ & rSVDdpd + DICMR & $0.40$ &  $1.37$ \\
             & $1$ & Decreasing & rSVDdpd + WCV & $0.32$ & $1.56$ & rSVDdpd + DICMR & $0.20$ &  $1.51$ \\  
             \midrule
            \multirow{3}{*}{$0.05$} & $0.05$ & Equal & rSVDdpd + WCV & $1.00$ & $0.00$ & rSVDdpd + IC3 & $0.92$ & $0.28$ \\
             & $0.5$ & Equal & rSVDdpd + WCV & $1.00$ & $0.00$ & rSVDdpd + DICMR & $0.57$ &  $1.04$ \\
             & $1$ & Equal & rSVDdpd + ECV & $0.70$ & $1.99$ & rSVDdpd + DICMR & $0.31$ &  $1.24$ \\  
             \midrule
            \multirow{3}{*}{$0.05$} & $0.05$ & Decreasing & rSVDdpd + WCV, ECV & $1.00$ & $0.00$ & rSVDdpd + DICMR & $0.90$ &  $0.32$ \\
             & $0.5$ & Decreasing & rSVDdpd + ECV & $0.89$ & $0.41$ & rSVDdpd + DICMR & $0.52$ &  $0.87$ \\
             & $1$ & Decreasing & rSVDdpd + WCV & $0.24$ & $1.88$ & rSVDdpd + DICMR & $0.22$ &  $1.87$ \\  
             \midrule
            \multirow{3}{*}{$0.1$} & $0.05$ & Equal & rSVDdpd + WCV, ECV & $1.00$ & $0.00$ & rSVDdpd + DICMR & $0.84$ & $0.40$  \\
             & $0.5$ & Equal & rSVDdpd + WCV & $1.00$ & $0.00$ & rSVDdpd + DICMR & $0.63$ &  $1.00$ \\
             & $1$ & Equal & rSVDdpd + WCV & $0.56$ & $1.17$ & rSVDdpd + DICMR & $0.40$ &  $1.22$ \\  
             \midrule
            \multirow{3}{*}{$0.1$} & $0.05$ & Decreasing & rSVDdpd + WCV, ECV & $1.00$ & $0.00$ & rSVDdpd + DICMR & $0.94$ &  $0.24$ \\
             & $0.5$ & Decreasing & rSVDdpd + ECV & $0.79$ & $1.11$ & rSVDdpd + DICMR & $0.52$ &  $0.84$  \\
             & $1$ & Decreasing & rSVDdpd + WCV & $0.22$ & $1.82$ & rSVDdpd + DICMR & $0.13$ &   $2.14$ \\  
             \midrule
            \multirow{3}{*}{$0.2$} & $0.05$ & Equal & rSVDdpd + WCV, ECV & $1.00$ & $0.00$ & rSVDdpd + DICMR & $0.88$ &  $0.35$ \\
             & $0.5$ & Equal & rSVDdpd + WCV & $0.99$ & $0.26$ & rSVDdpd + DICMR & $0.78$ &  $0.53$ \\
             & $1$ & Equal & rSVDdpd + WCV & $0.39$ & $1.30$ & rSVDdpd + DICMR & $0.09$ &  $2.95$ \\  
             \midrule
            \multirow{3}{*}{$0.2$} & $0.05$ & Decreasing & rSVDdpd + ECV & $1.00$ & $0.0$ & rSVDdpd + DICMR & $0.96$ &  $0.20$ \\
             & $0.5$ & Decreasing & rSVDdpd + WCV & $0.33$ & $1.17$ & rSVDdpd + DICMR & $0.52$ &  $0.93$ \\
             & $1$ & Decreasing & rSVDdpd + WCV & $0.26$ & $1.57$ & rSVDdpd + DICMR & $0.03$ &  $3.06$ \\  
            \bottomrule 
        \end{tabular}
    }
\end{table}

As evident from Table~\ref{tab:rank-result-summary}, the proposed DICMR is often the best performing penalized criterion, except in a few cases when there is no contamination. The Wold style cross-validation (WCV) method along with the robust rSVDdpd algorithm is usually the best for robust rank estimation problems which also take the longest amount of time. For a quicker solution, DICMR provides a reasonably good estimate that is often closely competitive to the WCV method, and is sometimes better than other cross-validation methods such as Gabriel-style cross-validation (GCV) or Bi-cross validation (BCV) methods.

We also note that when the errors are generated from a contaminated distribution, all of the existing methods (both penalized and CV approaches) that use traditional SVD fail to provide a reasonable estimate of the rank of the matrix, even when the contamination proportion is low. Even when the cross-validation methods are modified to use MAE or MAD as a scale measure for aggregating cross-validation errors, they do not produce a robust estimate of rank. This can be understood by the following phenomenon: If there is a single outlier in a sample of $n$ observations, then the usual $k$-fold cross-validation produces $(k-1)$ subsamples which contain the outlier. Hence, each of those $(k-1)$ subsamples out of total $k$ subsamples will produce an egregiously bad estimate of the parameter of interest, resulting in $100(1-1/k)\%$ outlying predictions. Because of this extremely high proportion of contamination, the final estimate will also be nonrobust, despite using any robust metric to obtain the final cross-validation measure; see the detailed results in Tables~\ref{tab:rank-result-setA-svd1}-\ref{tab:rank-result-setB-svd2} of the Appendix for illustrations.

When the robust rSVDdpd procedure is used instead of the traditional SVD, most of the rank estimation methods show improvements, except for the classical information criteria like AIC and BIC. Interestingly, even the simple elbow method using a robust estimate of singular values performs so well that it becomes competitive with some of the cross-validation methods. While we incorporate the rSVDdpd for most of the rank estimation procedures for comparison purposes, there is no obvious way to incorporate it into the Bayesian procedure of~\cite{hoff2007model}, so we refrain from considering that method in this scenario. The corresponding results are illustrated through Tables~\ref{tab:rank-result-setC-svd1}-\ref{tab:rank-result-setD-svd2} of the Appendix.

\section{An application in microarray data imputation}\label{sec:rank-real-application}

\begin{figure}[htpb]
    \centering
    \includegraphics[width=\linewidth]{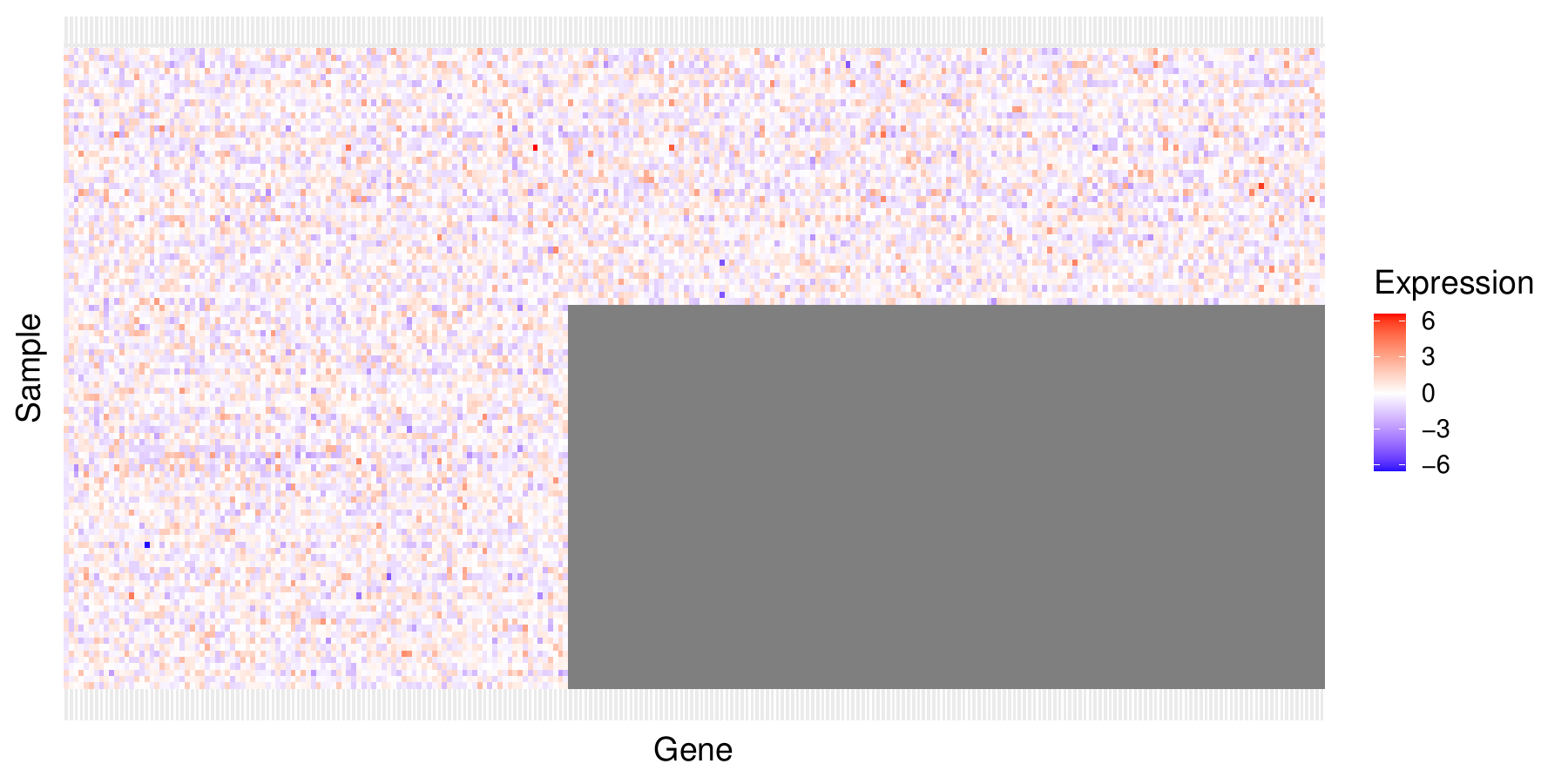}
    \caption{A partially observed PANCAN single-cell RNA sequencing dataset, with the gray shaded region denoting unavailable part of the data. Each column is preprocessed using log2 transformation and normalization.}
    \label{fig:gene-matrix-masked}
\end{figure}

Analysis of microarray data is an important tool in various biological and genetic studies. Microarray data records the expression profiles for various genomic markers obtained from various cells (e.g. by single cell RNA sequencing), which may belong to a single organism or multiple organisms. Such a dataset is usually described by a large high-dimensional data matrix, with the rows indicating various cells (or individuals) and the columns indicating different genes. Some notable challenges in analyzing these datasets are as follows:
\begin{enumerate}
    \item Single cell RNA-sequencing or access to commercially viable microarray services are costly, in terms of computational, storage and economical requirements.
    \item A significant number of entries in such a microarray data are usually zero resulting in a sparse data matrix. They may be biological zeros (i.e., the genetic marker is not expressed during sequencing) or technical zeros (i.e., the genetic marker was not measured during sequencing); see~\cite{ALRA_method} for further details. 
    \item Additionally, missing data appears if one combines various genome-wide studies to study correlations between various genes~\citep{cai2016structured}. For example, a large-scale sequencing study for a large corpus of genetic markers may be only available for a few cells (or individuals) while many small-scale studies may consider a larger set of individuals but restricted to a much smaller pool of genes. This can be viewed as a matrix completion exercise for a data matrix where a partition of the matrix is unobserved and needs to be estimated; see the gray region of Figure~\ref{fig:gene-matrix-masked}.
\end{enumerate}

As a result, there have been a growing interest in performing missing data imputation to recover the true counts of a microarray data. \cite{moorthy2019survey} provide a nice review of various traditional imputation methods for microarray datasets including K-nearest neighbour and SVD based algorithms. In addition, there has been a vast literature on matrix completion methods; see~\cite{mao2019matrix} and references therein. However, the zero-inflated distribution of the entries of the data matrix require more robust and specialized imputation methods compared to the nonrobust traditional techniques. Some notable algorithms are Markov Affinity-based Graph Imputation of Cells (MAGIC)~\citep{MAGIC_method}, single cell RNA sequencing impute (scImpute)~\cite{scImpute_method}, Single-cell Analysis via Expression Recovery (SAVER)~\citep{SAVER_method}, microbiome data imputation (mbImpute)~\citep{mbImpute_method} and Adaptively thresholded Low-Rank Approximation (ALRA)~\citep{ALRA_method}. Despite the existence of these specialized imputation techniques, there is no single algorithm that outperforms all others. \cite{Hou_Ji_Ji_Hicks_2020} performed a comprehensive benchmark study to compare the performances of these methods across different real-life and simulated datasets, and concluded that one of the main challenge of imputing gene expression values is the presence of a few abnormal counts in diseased cells. Clearly, a robust imputation procedure unaffected by these few outlying measurements would be able to provide a more reasonable estimate of the gene expression profiles.

To illustrate the usefulness of the proposed robust rank estimation technique, we consider an RNA sequencing dataset of various cancer cells~\cite{gene_expression_data} available at The Cancer Genome Atlas (TCGA). The dataset contains gene expressions in log2 scale for $801$ tumor cells and $20531$ genes from various patients suffering from one of five types of cancer. Given this entire dataset $\bb{X}$, we partition it as
\begin{equation*}
    \bb{P}_1\bb{X}\bb{P}_2 = \begin{bmatrix}
        \bb{X}_{11} & \bb{X}_{12}\\
        \bb{X}_{21} & \bb{X}_{22}
    \end{bmatrix},
\end{equation*}
\noindent where $\bb{X}_{11}$ is a matrix of dimension $100 \times 2000$, $\bb{P}_1$ and $\bb{P}_2$ are random permutation matrices of respective orders. Then, we remove the entries $\bb{X}_{22}$ and mark them as missing data. This results in a data matrix where $10\%$ of the patients (rows) have complete data on all genes, while only for $10\%$ of the genes (columns), expression profiles are available across all patients. Assuming the LSN decomposition~\eqref{eqn:rank-lsn-decomp} for this data matrix, we apply our DICMR algorithm to estimate the rank. Given the estimated rank $\widehat{r}$, the estimates of the missing observations is then obtained as
\begin{equation*}
    \bbhat{X}_{22} = \bbhat{L}^{(\widehat{r})}_{\alpha, 21} \left( \bbhat{L}^{(\widehat{r})}_{\alpha, 11} \right)^{+} \bbhat{L}^{(\widehat{r})}_{\alpha, 12}, 
\end{equation*}
\noindent where $\bbhat{L}_{\alpha,ij}^{(r)}$ is the estimate of the low-rank component using rSVDdpd estimates upto rank $r$ for the block $\bb{X}_{ij}$, $i = 1, 2$, $j = 1, 2$ with robustness parameter $\alpha$, and $\bb{A}^+$ is the Moore-Penrose generalized inverse of the matrix $\bb{A}$. We also consider a version of the imputation procedure where before performing the imputation, a normalization step is performed on each column of the observed data matrix $\bb{X}$ as, $Z_{ij} = (X_{ij} - \mu_i) / \sigma_i$ for each $i = 1, \dots, n; j = 1, \dots, p$. Here, we take $\mu_i$ and $\sigma_i$ to be the median and the median absolute deviation about median of the observed values of $i$-th column of $\bb{X}$ matrix. The imputation is then performed on the normalized matrix $\bb{Z}$ and imputed matrix was then rescaled back to provide imputed gene expression counts. 

\begin{table}[htbp]
    \centering
    \resizebox*{\textwidth}{!}{\begin{tabular}{lr}
        \toprule
        \textbf{Method} & \textbf{Relative RMSE}\\
        \midrule
        knnImpute~\citep{knnImpute_method} & $0.0365$\\
        llsImpute~\citep{llsImpute_method} & $0.0272$\\
        scImpute~\citep{scImpute_method} & $0.7497$\\
        MAGIC~\citep{MAGIC_method} & $0.0881$\\
        SAVER without normalization~\citep{SAVER_method} & $0.7508$\\
        SAVER with normalization~\citep{SAVER_method} & $0.0259$\\
        ALRA~\citep{ALRA_method} & $0.7468$\\
        rSVDdpd without normalization; $\hat{\alpha} = 0.458, \hat{r} = 30$~\citep{roy2024robustsvd} & $0.4564$\\
        rSVDdpd with normalization; $\hat{\alpha} = 0.875, \hat{r} = 5$~\citep{roy2024robustsvd} & $\mathbf{0.0237}$\\
        \bottomrule
    \end{tabular}}
    \caption{The relative RMSE of various imputation methods for the matrix completion problem on RNA-seq PANCAN dataset.}
    \label{tab:microarray-compare}
\end{table}

\begin{figure}[htbp]
    \centering
    \includegraphics[width = \textwidth]{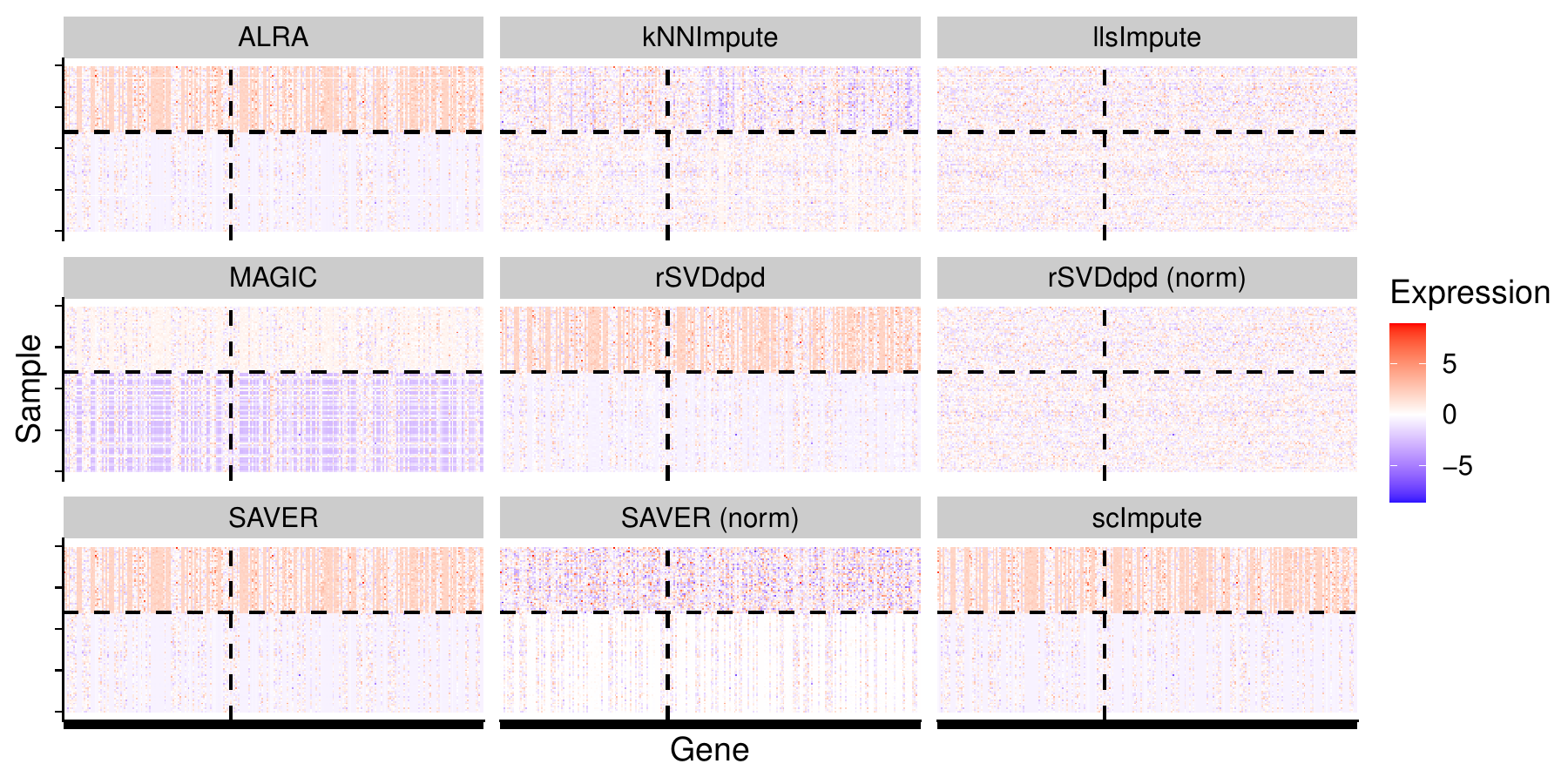}
    \caption{The imputed RNA-seq PANCAN dataset obtained by different imputation techniques; the dashed line shows the partition for which the bottom-right corner block is imputed; ``norm'' in brackets indicate a preprocssing normalization step is performed.}
    \label{fig:pancan-imputed-methods}
\end{figure}

To compare the performance of various imputation techniques, we track the relative RMSE metric given by
\begin{equation*}
    \text{Relative RMSE} = \dfrac{\Vert \bb{X}_{22} - \bbhat{X}_{22}\Vert_F^2}{\Vert \bb{X}_{22}\Vert_F^2}    
\end{equation*}
\noindent where $\Vert\bb{A}\Vert_F^2$ denotes the squared Frobenius norm of a matrix $\bb{A}$. The relative RMSE obtained are illustrated in Table~\ref{tab:microarray-compare}, we illustrate the performance of various imputation methods. For the rSVDdpd algorithm, the choice of the robustness procedure is obtained based on a conditional MSE criterion as illustrated in~\cite{roy2024robustsvd} and the rank of the data matrix is estimated using the proposed DICMR criterion described in Section~\ref{sec:rank-dicmr}. It is quite clear from Table~\ref{tab:microarray-compare} that the normalization step significantly improves the quality of imputation. With this normalization preprocessing, rSVDdpd estimator is able to achieve the best performance among all the methods compared here. The closest contenders are the SAVER algorithm with a normalization step, llsImpute and MAGIC methods both of which perform a normalization step internally. In Figure~\ref{fig:pancan-imputed-methods}, we illustrate the imputed microarray data as obtained by various methods. All the algorithms, without a normalization step underestimates the entries of the matrix. Based on a subjective evaluation, the llsImpute and the rSVDdpd with a normalization step completes the matrix in a way that is coherent with the observed patterns. The dataset~\citep{gene_expression_data} also contains a label containing the type of the cancer (which we do not make use of) for classification purposes. It includes $5$ types of cancers, namely BRCA (Breast invasive carcinoma), COAD (Colon adenocarcinoma), KIRC (Kidney renal clear cell carcinoma), LUAD (Lung adenocarcinoma) and PRAD (Prostate adenocarcinoma), which surprising matches the estimate of the rank obtained by the DICMR criterion.

\begin{figure}[htpb]
    \centering
    \includegraphics[width=\linewidth]{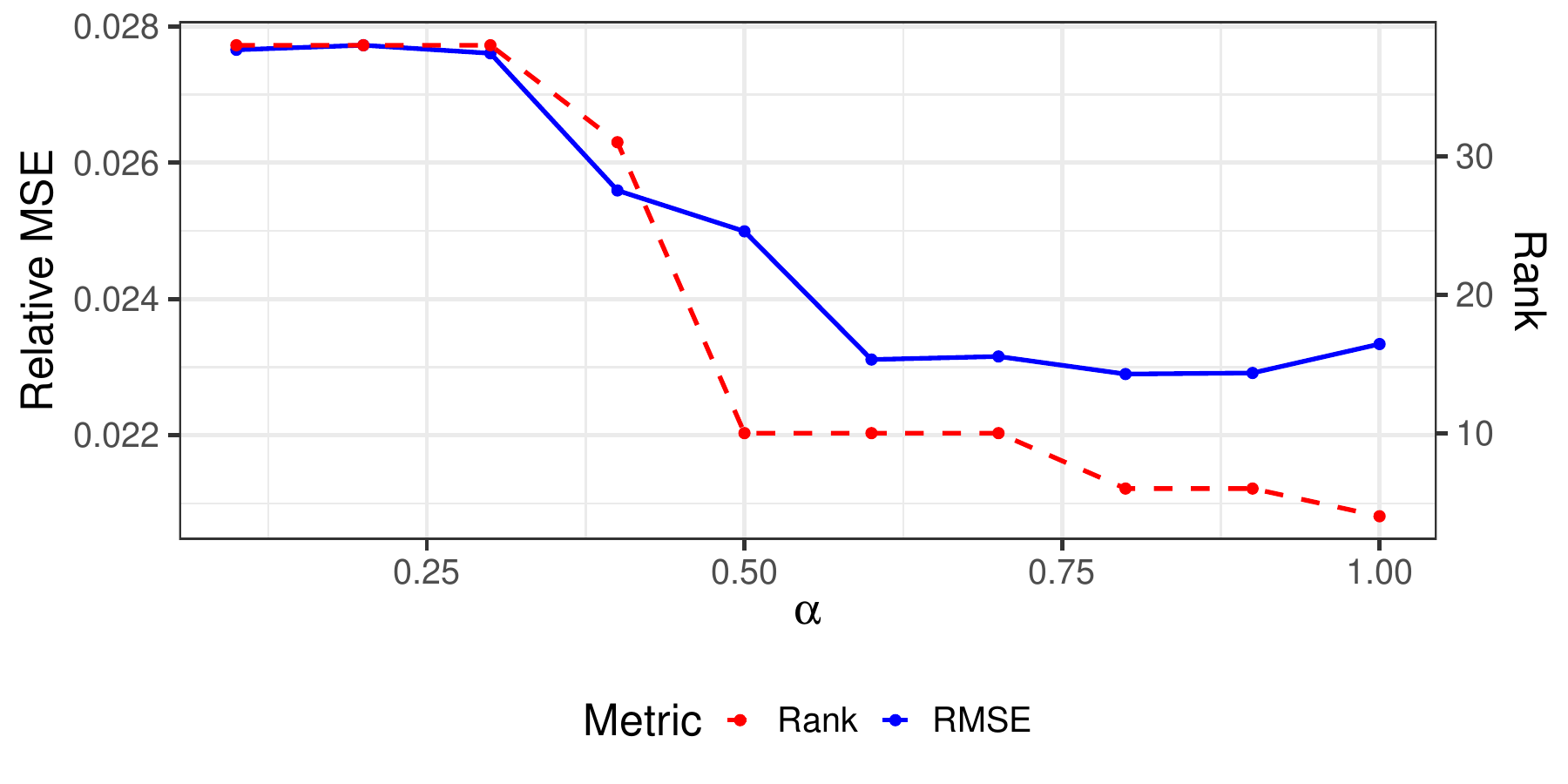}
    \caption{Estimated rank and relative MSE based on DICMR and rSVDdpd algorithm}
    \label{fig:rank-metric-alpha}
\end{figure}

Since DICMR is a robust rank estimation criterion, the estimated rank should monotonically change as the robustness parameter $\alpha$ increases. In Figure~\ref{fig:rank-metric-alpha}, we illustrate this phenomenon through a monitoring plot; see \cite{monitoring_book} for a definition and purpose of such plots. It is also worthwhile to note that the chosen optimal $\alpha = 0.825$ based on the conditional MSE criterion as in~\cite{roy2024robustsvd}, closely corresponds to the region where the actual relative MSE is minimized. The sharp decrease around $\alpha = 0.5$ for the estimated rank suggest that the there is a significant presence of outlying observations in the dataset, and the DICMR algorithm provides a significantly different estimate of the rank compared to the classical nonrobust SVD based methods.

\section{Conclusion}

The current matrix rank estimation methods either achieve poor robustness or become computationally infeasible for practical applications when equipped with standard robust matrix factorization techniques. This paper addresses this gap by proposing a novel and computationally simple rank estimation technique based on an information criterion DICMR with provable robustness guarantees. Our theoretical analysis establishes that DICMR is B-robust, ensuring stability of rank estimation even under significant contamination or model misspecification. We also establish that DICMR never asymptotically underestimates the true rank, and provide an upper bound on the probability of overestimation. Through extensive simulations, DICMR consistently outperformed existing penalized methods and remained competitive with computationally intensive cross-validation approaches, while maintaining substantially lower computational costs. The empirical studies also demonstrated that DICMR maintains reliable performance across varying noise-to-signal ratios and contamination levels. Even when there are no outliers, if the noise-to-signal ratio is high, using the DICMR criterion leads to better performances than traditional information criteria. As a practical demonstration, we analyzed imputation techniques for RNA-seq microarray data. By using a combination of an appropriate normalization step, the robust rSVDdpd procedure, and the proposed method DICMR yielded the lowest reconstruction error among the state-of-the-art techniques. This result underscores DICMR's potential as a simple and useful tool for high-dimensional biological data analysis. 

Despite its strong theoretical and empirical performances, DICMR may overestimate the true rank, when the data matrix is nearly square. In many contexts, this may not be detrimental. In contrast to underestimation, which irreversibly discards any valid statistical signal present in the matrix data, a slight overestimation may add some redundant components that can be subsequently pruned by other methods. We believe, DICMR can serve as a quick and fast, but dependable upper bound estimator for the rank of a data matrix. A promising future direction could be to develop a hybrid approach that uses DICMR to reduce the search space to a set of feasible candidate ranks, and subsequently uses a more extensive cross-validation method on a robust estimate (using rSVDdpd) of the data matrix for a more exact search. A viable approach for the non-robust case is available in~\cite{barigozzi2020consistent}. Another interesting future direction would be to extend the criterion for estimating tubal-ranks of tensors and exploring its applications beyond genetics.

\bibliography{references}

\begin{thebibliography}{45}
\providecommand{\natexlab}[1]{#1}
\providecommand{\url}[1]{\texttt{#1}}
\expandafter\ifx\csname urlstyle\endcsname\relax
  \providecommand{\doi}[1]{doi: #1}\else
  \providecommand{\doi}{doi: \begingroup \urlstyle{rm}\Url}\fi

\bibitem[Akaike(1973)]{akaike1973information}
Hirotogu Akaike.
\newblock \emph{Information Theory and an Extension of the Maximum Likelihood Principle}, pages 199--213.
\newblock Springer New York, New York, NY, 1973.

\bibitem[Arciniegas-Alarc{\'o}n et~al.(2022)Arciniegas-Alarc{\'o}n, García-Peña, and Krzanowski]{alarcon2022cv}
Sergio Arciniegas-Alarc{\'o}n, Marisol García-Peña, and Wojtek~J. Krzanowski.
\newblock {Cross-Validation for Lower Rank Matrices Containing Outliers}.
\newblock \emph{Applied System Innovation}, 5\penalty0 (4), 2022.
\newblock ISSN 2571-5577.
\newblock \doi{10.3390/asi5040069}.

\bibitem[Atkinson et~al.(2025)Atkinson, Riani, Corbellini, Perrotta, and Todorov]{monitoring_book}
Anthony~C. Atkinson, Marco Riani, Aldo Corbellini, Domenico Perrotta, and Valentin Todorov.
\newblock \emph{{Robust statistics through the monitoring approach: Applications in regression}}.
\newblock Springer series in statistics. Springer International Publishing, Cham, Switzerland, July 2025.
\newblock ISBN 9783031883644.
\newblock \doi{10.1007/978-3-031-88365-1}.

\bibitem[Bai and Ng(2002)]{bai2002determining}
Jushan Bai and Serena Ng.
\newblock {Determining the Number of Factors in Approximate Factor Models}.
\newblock \emph{Econometrica}, 70\penalty0 (1):\penalty0 191--221, 2002.
\newblock \doi{https://doi.org/10.1111/1468-0262.00273}.

\bibitem[Barigozzi and Cho(2020)]{barigozzi2020consistent}
Matteo Barigozzi and Haeran Cho.
\newblock {Consistent estimation of high-dimensional factor models when the factor number is over-estimated}.
\newblock \emph{Electronic Journal of Statistics}, 14\penalty0 (2):\penalty0 2892 -- 2921, 2020.
\newblock \doi{10.1214/20-EJS1741}.

\bibitem[Basu et~al.(1998)Basu, Harris, Hjort, and Jones]{basu1998robust}
Ayanendranath Basu, Ian~R. Harris, Nils~L. Hjort, and M.~C. Jones.
\newblock {Robust and Efficient Estimation by Minimising a Density Power Divergence}.
\newblock \emph{Biometrika}, 85\penalty0 (3):\penalty0 549--559, 1998.
\newblock ISSN 00063444.

\bibitem[Bro et~al.(2008)Bro, Kjeldahl, Smilde, and Kiers]{bro2008cv}
R.~Bro, K.~Kjeldahl, A.~K. Smilde, and H.~A.~L. Kiers.
\newblock {Cross-validation of component models: A critical look at current methods}.
\newblock \emph{Analytical and Bioanalytical Chemistry}, 390\penalty0 (5):\penalty0 1241--1251, Mar 2008.
\newblock ISSN 1618-2650.
\newblock \doi{10.1007/s00216-007-1790-1}.

\bibitem[Cai et~al.(2016)Cai, Cai, and Zhang]{cai2016structured}
Tianxi Cai, T~Tony Cai, and Anru Zhang.
\newblock {Structured matrix completion with applications to genomic data integration}.
\newblock \emph{Journal of the American Statistical Association}, 111\penalty0 (514):\penalty0 621--633, 2016.

\bibitem[Cand\`{e}s et~al.(2011)Cand\`{e}s, Li, Ma, and Wright]{candes2011robust}
Emmanuel~J. Cand\`{e}s, Xiaodong Li, Yi~Ma, and John Wright.
\newblock {Robust Principal Component Analysis?}
\newblock \emph{J. ACM}, 58\penalty0 (3), 6 2011.
\newblock ISSN 0004-5411.
\newblock \doi{10.1145/1970392.1970395}.

\bibitem[Choi et~al.(2025)Choi, Kwon, and Liao]{choi2025inference}
Jungjun Choi, Hyukjun Kwon, and Yuan Liao.
\newblock {Inference for Low-rank Models without Estimating the Rank}.
\newblock \emph{Journal of the American Statistical Association}, 0\penalty0 (ja):\penalty0 1--22, 2025.
\newblock \doi{10.1080/01621459.2025.2538272}.

\bibitem[Choi et~al.(2017)Choi, Taylor, and Tibshirani]{choi2017selecting}
Yunjin Choi, Jonathan Taylor, and Robert Tibshirani.
\newblock {Selecting the number of principal components: Estimation of the true rank of a noisy matrix}.
\newblock \emph{{The Annals of Statistics}}, pages 2590--2617, 2017.

\bibitem[Cichocki et~al.(2011)Cichocki, Cruces, and Amari]{chicoki2011robustnmf}
Andrzej Cichocki, Sergio Cruces, and Shun-ichi Amari.
\newblock {Generalized Alpha-Beta Divergences and Their Application to Robust Nonnegative Matrix Factorization}.
\newblock \emph{Entropy}, 13\penalty0 (1):\penalty0 134--170, 2011.
\newblock ISSN 1099-4300.
\newblock \doi{10.3390/e13010134}.

\bibitem[Du et~al.(2017)Du, Ma, Li, and Ma]{du2017robust}
Shiqiang Du, Yide Ma, Shouliang Li, and Yurun Ma.
\newblock {Robust unsupervised feature selection via matrix factorization}.
\newblock \emph{Neurocomputing}, 241:\penalty0 115--127, 2017.
\newblock ISSN 0925-2312.
\newblock \doi{https://doi.org/10.1016/j.neucom.2017.02.034}.

\bibitem[Eastment and Krzanowski(1982)]{eastment1982crossvalidation}
H.~T. Eastment and W.~J. Krzanowski.
\newblock Cross-{Validatory} {Choice} of the {Number} of {Components} from a {Principal} {Component} {Analysis}.
\newblock \emph{Technometrics}, 24\penalty0 (1):\penalty0 73--77, 1982.
\newblock ISSN 0040-1706.
\newblock \doi{10.2307/1267581}.

\bibitem[Fiorini(2016)]{gene_expression_data}
Samuele Fiorini.
\newblock {Gene Expression Cancer RNA-Seq}.
\newblock UCI Machine Learning Repository, 2016.

\bibitem[Gabriel(2002)]{gabriel2002lebiplot}
K.~Ruben Gabriel.
\newblock {Le biplot - outil d'exploration de données multidimensionnelles}.
\newblock \emph{Journal de la Société française de statistique}, 143\penalty0 (3-4):\penalty0 5--55, 2002.
\newblock ISSN 1625-7421.

\bibitem[Ghosh and Basu(2013)]{ghosh2013robust}
Abhik Ghosh and Ayanendranath Basu.
\newblock {Robust estimation for independent non-homogeneous observations using density power divergence with applications to linear regression}.
\newblock \emph{Electronic Journal of Statistics}, 7\penalty0 (none):\penalty0 2420 -- 2456, 2013.
\newblock \doi{10.1214/13-EJS847}.

\bibitem[Greenacre et~al.(2022)Greenacre, Groenen, Hastie, D'Enza, Markos, and Tuzhilina]{greenacre2022principal}
Michael Greenacre, Patrick J.~F. Groenen, Trevor Hastie, Alfonso~Iodice D'Enza, Angelos Markos, and Elena Tuzhilina.
\newblock {Principal component analysis}.
\newblock \emph{Nature Reviews Methods Primers}, 2\penalty0 (1):\penalty0 100, Dec 2022.
\newblock ISSN 2662-8449.
\newblock \doi{10.1038/s43586-022-00184-w}.

\bibitem[Hoff(2007)]{hoff2007model}
Peter~D. Hoff.
\newblock Model {Averaging} and {Dimension} {Selection} for the {Singular} {Value} {Decomposition}.
\newblock \emph{Journal of the American Statistical Association}, 102\penalty0 (478):\penalty0 674--685, 2007.
\newblock ISSN 0162-1459.

\bibitem[Hou et~al.(2020)Hou, Ji, Ji, and Hicks]{Hou_Ji_Ji_Hicks_2020}
Wenpin Hou, Zhicheng Ji, Hongkai Ji, and Stephanie~C. Hicks.
\newblock {A systematic evaluation of single-cell RNA-sequencing imputation methods}.
\newblock \emph{Genome biology}, 21\penalty0 (1):\penalty0 218, August 2020.
\newblock ISSN 1474-7596.
\newblock \doi{10.1186/s13059-020-02132-x}.

\bibitem[Huang et~al.(2018)Huang, Wang, Torre, Dueck, Shaffer, Bonasio, Murray, Raj, Li, and Zhang]{SAVER_method}
Mo~Huang, Jingshu Wang, Eduardo Torre, Hannah Dueck, Sydney Shaffer, Roberto Bonasio, John~I. Murray, Arjun Raj, Mingyao Li, and Nancy~R. Zhang.
\newblock {SAVER: gene expression recovery for single-cell RNA sequencing}.
\newblock \emph{Nature methods}, 15\penalty0 (7):\penalty0 539--542, July 2018.
\newblock ISSN 1548-7091.
\newblock \doi{10.1038/s41592-018-0033-z}.

\bibitem[Jiang et~al.(2021)Jiang, Li, and Li]{mbImpute_method}
Ruochen Jiang, Wei~Vivian Li, and Jingyi~Jessica Li.
\newblock {mbImpute: an accurate and robust imputation method for microbiome data}.
\newblock \emph{Genome biology}, 22\penalty0 (1):\penalty0 192, June 2021.
\newblock ISSN 1474-7596.
\newblock \doi{10.1186/s13059-021-02400-4}.

\bibitem[Jolliffe(2002)]{jolliffe_principal_2002}
Ian~T Jolliffe.
\newblock \emph{Principal {Component} {Analysis}}.
\newblock Springer {Series} in {Statistics}. Springer-Verlag, New York, 2002.
\newblock ISBN 978-0-387-95442-4.
\newblock \doi{10.1007/b98835}.

\bibitem[Karagrigoriou and Papaioannou(2008)]{karagrigoriou2008measures}
Alex Karagrigoriou and Takis Papaioannou.
\newblock {On measures of information and divergence and model selection criteria}.
\newblock \emph{Statistical Models and Methods for Biomedical and Technical Systems}, pages 503--518, 2008.

\bibitem[Kim et~al.(2005)Kim, Golub, and Park]{llsImpute_method}
Hyunsoo Kim, Gene~H. Golub, and Haesun Park.
\newblock {Missing value estimation for DNA microarray gene expression data: local least squares imputation}.
\newblock \emph{Bioinformatics (Oxford, England)}, 21\penalty0 (2):\penalty0 187–198, January 2005.
\newblock ISSN 1367-4803.
\newblock \doi{10.1093/bioinformatics/bth499}.

\bibitem[Kurata(2022)]{kurata2022robustness}
Sumito Kurata.
\newblock {On robustness of model selection criteria based on divergence measures: Generalizations of BHHJ divergence-based method and comparison}.
\newblock \emph{Communications in Statistics - Theory and Methods}, 0\penalty0 (0):\penalty0 1--18, 2022.
\newblock \doi{10.1080/03610926.2022.2155788}.

\bibitem[Kurata and Hamada(2018)]{kurata2018robust}
Sumito Kurata and Etsuo Hamada.
\newblock {A robust generalization and asymptotic properties of the model selection criterion family}.
\newblock \emph{Communications in Statistics - Theory and Methods}, 47\penalty0 (3):\penalty0 532--547, 2018.
\newblock \doi{10.1080/03610926.2017.1307405}.

\bibitem[Kurata and Hamada(2020)]{kurata2020consistency}
Sumito Kurata and Etsuo Hamada.
\newblock On the consistency and the robustness in model selection criteria.
\newblock \emph{Communications in Statistics - Theory and Methods}, 49\penalty0 (21):\penalty0 5175--5195, 2020.
\newblock \doi{10.1080/03610926.2019.1615093}.

\bibitem[Li and Li(2018)]{scImpute_method}
Wei~Vivian Li and Jingyi~Jessica Li.
\newblock An accurate and robust imputation method scimpute for single-cell rna-seq data.
\newblock \emph{Nature communications}, 9\penalty0 (1):\penalty0 997, March 2018.
\newblock ISSN 2041-1723.
\newblock \doi{10.1038/s41467-018-03405-7}.

\bibitem[Linderman et~al.(2022)Linderman, Zhao, Roulis, Bielecki, Flavell, Nadler, and Kluger]{ALRA_method}
George~C. Linderman, Jun Zhao, Manolis Roulis, Piotr Bielecki, Richard~A. Flavell, Boaz Nadler, and Yuval Kluger.
\newblock {Zero-preserving imputation of single-cell RNA-seq data}.
\newblock \emph{Nature communications}, 13\penalty0 (1):\penalty0 192, January 2022.
\newblock ISSN 2041-1723.
\newblock \doi{10.1038/s41467-021-27729-z}.

\bibitem[Machado(1993)]{machado1993robust}
Jose~AF Machado.
\newblock Robust model selection and m-estimation.
\newblock \emph{Econometric Theory}, 9\penalty0 (3):\penalty0 478--493, 1993.

\bibitem[Mao et~al.(2019)Mao, Chen, and Wong]{mao2019matrix}
Xiaojun Mao, Song~Xi Chen, and Raymond K.~W. Wong.
\newblock Matrix completion with covariate information.
\newblock \emph{Journal of the American Statistical Association}, 114\penalty0 (525):\penalty0 198--210, 2019.
\newblock \doi{10.1080/01621459.2017.1389740}.

\bibitem[Moorthy et~al.(2019)Moorthy, Jaber, Ismail, Ernawan, Mohamad, and Deris]{moorthy2019survey}
Kohbalan Moorthy, Aws~Naser Jaber, Mohd~Arfian Ismail, Ferda Ernawan, Mohd~Saberi Mohamad, and Safaai Deris.
\newblock \emph{Missing-Values Imputation Algorithms for Microarray Gene Expression Data}, pages 255--266.
\newblock Springer, New York, NY, 2019.
\newblock ISBN 9781493994427.
\newblock \doi{10.1007/978-1-4939-9442-7_12}.

\bibitem[Owen and Perry(2009)]{owen2009bi}
Art~B. Owen and Patrick~O. Perry.
\newblock {Bi-cross-validation of the {SVD} and the nonnegative matrix factorization}.
\newblock \emph{The Annals of Applied Statistics}, 3\penalty0 (2):\penalty0 564--594, June 2009.
\newblock ISSN 1932-6157, 1941-7330.
\newblock \doi{10.1214/08-AOAS227}.
\newblock Publisher: Institute of Mathematical Statistics.

\bibitem[Roy et~al.(2024{\natexlab{a}})Roy, Basu, and Ghosh]{roy2024robustpca}
Subhrajyoty Roy, Ayanendranath Basu, and Abhik Ghosh.
\newblock Robust principal component analysis using density power divergence.
\newblock \emph{Journal of Machine Learning Research}, 25\penalty0 (324):\penalty0 1--40, 2024{\natexlab{a}}.

\bibitem[Roy et~al.(2024{\natexlab{b}})Roy, Ghosh, and Basu]{roy2024robustsvd}
Subhrajyoty Roy, Abhik Ghosh, and Ayanendranath Basu.
\newblock Robust singular value decomposition with application to video surveillance background modelling.
\newblock \emph{Statistics and Computing}, 34\penalty0 (5):\penalty0 178, 2024{\natexlab{b}}.

\bibitem[Sadek(2012)]{sadek2012svd}
Rowayda~A Sadek.
\newblock {SVD based image processing applications: state of the art, contributions and research challenges}.
\newblock \emph{arXiv preprint arXiv:1211.7102}, 2012.

\bibitem[Schwarz(1978)]{schwarz1978estimating}
Gideon Schwarz.
\newblock {Estimating the Dimension of a Model}.
\newblock \emph{The Annals of Statistics}, 6\penalty0 (2):\penalty0 461 -- 464, 1978.
\newblock \doi{10.1214/aos/1176344136}.

\bibitem[Shabalin and Nobel(2013)]{shabalin2013reconstruction}
Andrey~A. Shabalin and Andrew~B. Nobel.
\newblock {Reconstruction of a low-rank matrix in the presence of Gaussian noise}.
\newblock \emph{Journal of Multivariate Analysis}, 118:\penalty0 67--76, 2013.
\newblock ISSN 0047-259X.
\newblock \doi{https://doi.org/10.1016/j.jmva.2013.03.005}.

\bibitem[Thorndike(1953)]{thorndike1953elbow}
Robert~L. Thorndike.
\newblock {Who belongs in the family?}
\newblock \emph{Psychometrika}, 18\penalty0 (4):\penalty0 267--276, Dec 1953.
\newblock ISSN 1860-0980.
\newblock \doi{10.1007/BF02289263}.

\bibitem[Toma et~al.(2020)Toma, Karagrigoriou, and Trentou]{toma2020robustmodel}
Aida Toma, Alex Karagrigoriou, and Paschalini Trentou.
\newblock {Robust Model Selection Criteria Based on Pseudodistances}.
\newblock \emph{Entropy}, 22\penalty0 (3), 2020.
\newblock ISSN 1099-4300.
\newblock \doi{10.3390/e22030304}.

\bibitem[Troyanskaya et~al.(2001)Troyanskaya, Cantor, Sherlock, Brown, Hastie, Tibshirani, Botstein, and Altman]{knnImpute_method}
O.~Troyanskaya, M.~Cantor, G.~Sherlock, P.~Brown, T.~Hastie, R.~Tibshirani, D.~Botstein, and R.~B. Altman.
\newblock {Missing value estimation methods for DNA microarrays}.
\newblock \emph{Bioinformatics (Oxford, England)}, 17\penalty0 (6):\penalty0 520--525, June 2001.
\newblock ISSN 1367-4803.
\newblock \doi{10.1093/bioinformatics/17.6.520}.

\bibitem[van Dijk et~al.(2018)van Dijk, Sharma, Nainys, Yim, Kathail, Carr, Burdziak, Moon, Chaffer, Pattabiraman, Bierie, Mazutis, Wolf, Krishnaswamy, and Pe'er]{MAGIC_method}
David van Dijk, Roshan Sharma, Juozas Nainys, Kristina Yim, Pooja Kathail, Ambrose~J. Carr, Cassandra Burdziak, Kevin~R. Moon, Christine~L. Chaffer, Diwakar Pattabiraman, Brian Bierie, Linas Mazutis, Guy Wolf, Smita Krishnaswamy, and Dana Pe'er.
\newblock {Recovering gene interactions from single-cell data using data diffusion}.
\newblock \emph{Cell}, 174\penalty0 (3):\penalty0 716--729.e27, July 2018.
\newblock ISSN 0092-8674.
\newblock \doi{10.1016/j.cell.2018.05.061}.

\bibitem[Wold(1978)]{wold1978crossvalidation}
Svante Wold.
\newblock Cross-{Validatory} {Estimation} of the {Number} of {Components} in {Factor} and {Principal} {Components} {Models}.
\newblock \emph{Technometrics}, 20\penalty0 (4):\penalty0 397--405, November 1978.
\newblock ISSN 0040-1706.
\newblock \doi{10.1080/00401706.1978.10489693}.

\bibitem[Xu et~al.(2021)Xu, He, Xie, and Wu]{xu2021adaptive}
Zhengqin Xu, Rui He, Shoulie Xie, and Shiqian Wu.
\newblock {Adaptive rank estimate in robust principal component analysis}.
\newblock In \emph{Proceedings of the IEEE/CVF Conference on Computer Vision and Pattern Recognition}, pages 6577--6586, 2021.

\end{thebibliography}

\pagebreak
\appendix

\section{Appendix: Proofs of the Results}\label{appendix:rank-proof}

\subsection{Proof of Theorem~\ref{thm:rank-selection-consistency-one}}\label{appendix:rank-proof-one}

Let us denote the penalty function of the DICMR as in~\eqref{eqn:rank-dicmr} as $r\Pcal_\alpha$. Now, consider the difference in the DICMR between rank zero and rank one,
\begin{align}
    & \text{DICMR}_\alpha(0) - \text{DICMR}_\alpha(1) \nonumber \\
    = \quad & H_\alpha^{(0)}(\bbhat{\theta}_\alpha^{(0)}; \bb{X}) - H_\alpha^{(1)}(\bbhat{\theta}_\alpha^{(1)}; \bb{X}) - \Pcal_\alpha \nonumber \\
    = \quad & \left( (\widehat{\sigma}_\alpha^{(0)})^{-\alpha} - (\widehat{\sigma}_\alpha^{(1)})^{-\alpha} \right) \norm{f}_{1+\alpha}^{1+\alpha} \nonumber \\
    & \quad + \left( 1 + \dfrac{1}{\alpha} \right) \dfrac{1}{np}\sum_{i=1}^n \sum_{j=1}^p \left[ (\widehat{\sigma}_\alpha^{(1)})^{-\alpha} f^\alpha\left( \dfrac{\vert X_{ij} - \widehat{\lambda}_{1}\widehat{u}_{1i}\widehat{v}_{1j} \vert}{\widehat{\sigma}_\alpha^{(1)}} \right)  - (\widehat{\sigma}_\alpha^{(0)})^{-\alpha} f^\alpha\left( \dfrac{\vert X_{ij} \vert}{\widehat{\sigma}_\alpha^{(0)}} \right) \right] - \Pcal_\alpha
    \label{eqn:rank-dicmr-difference}
\end{align}
\noindent where $(\widehat{\sigma}_\alpha^{(0)})^2$ and $(\widehat{\sigma}_\alpha^{(1)})^2$ are the estimated noise variance for rank zero and rank one robust SVD estimates. Similarly, let $\sigma_0^2$ and $\sigma_1^2$ be the corresponding population variance based on rank zero and rank one partial robust SVD approximation.

It has been shown in~\cite{roy2024robustsvd} that under Assumptions (B1)-(B6), the ``rSVDdpd'' estimates are asymptotically consistent under the regime when $n, p \to \infty$ such that $n / p \to c \in (0, \infty)$; see Theorem 7 of~\cite{roy2024robustsvd}. Therefore, we have $\widehat{\lambda}_1 \xrightarrow{P} \lambda_1$, $\Vert \widehat{\bb{u}}_{1} - \bb{u}_1\Vert \xrightarrow{P} 0$ and $\Vert \widehat{\bb{v}}_1 - \bb{v}_1\Vert \xrightarrow{P} 0$, where the symbol $\xrightarrow{P}$ indicates convergence in probability. Therefore, by a generalized version of Khinchin's law of large numbers, we have
\begin{equation*}
    \dfrac{1}{np}\sum_{i=1}^n \sum_{j=1}^p (\widehat{\sigma}_\alpha^{(1)})^{-\alpha} f^\alpha\left( \dfrac{\vert X_{ij} - \widehat{\lambda}_1 \widehat{u}_{1i}\widehat{v}_{1j} \vert }{\widehat{\sigma}_\alpha^{(1)}} \right) = \sigma_1^{-\alpha} \norm{f}_{1+\alpha}^{1+\alpha} + \Ocal(1/\sqrt{np}).
\end{equation*}
\noindent Similarly, we also have 
\begin{align*}
    & \dfrac{1}{np}\sum_{i=1}^n \sum_{j=1}^p (\widehat{\sigma}_\alpha^{(0)})^{-\alpha} f^\alpha\left( \dfrac{\vert X_{ij} \vert }{\widehat{\sigma}_\alpha^{(0)}} \right)\\
    = \quad & \dfrac{1}{np}\sum_{i=1}^n \sum_{j=1}^p \int (\widehat{\sigma}_\alpha^{(0)})^{-\alpha} \sigma_1^{-1} f^\alpha\left( \dfrac{\vert x\vert}{ \widehat{\sigma}_\alpha^{(0)} } \right) f\left( \dfrac{\vert x - \lambda_1 u_{1i}v_{1j} \vert}{\sigma_1} \right)dx + \Ocal(1/\sqrt{np})\\
    = \quad & \dfrac{1}{np}\sum_{i=1}^n \sum_{j=1}^p \int (\widehat{\sigma}_\alpha^{(0)})^{-\alpha} f^\alpha\left( \left\vert \widetilde{r}_\sigma z + \dfrac{\lambda_1 u_{1i} v_{1j}}{\widehat{\sigma}_\alpha^{(0)}} \right\vert \right)f\left( \vert z\vert \right)dz + \Ocal(1/\sqrt{np})
\end{align*}
\noindent where $\widetilde{r}_\sigma = \sigma_1 / \widehat{\sigma}_\alpha^{(0)}$. Since, $f$ is decreasing in the magnitude of its argument, symmetric about $0$ and it is a density function, it follows that 
\begin{equation*}
    \int f^\alpha\left( \left\vert \widetilde{r}_\sigma z + \dfrac{\lambda_1 u_{1i} v_{1j}}{\widehat{\sigma}_\alpha^{(0)}} \right\vert \right)f\left( \vert z\vert \right)dz \leq f^\alpha(0) \int f(\vert z\vert)dz = f^\alpha(0).
\end{equation*}
\noindent Now, putting all these back in~\eqref{eqn:rank-dicmr-difference}, we obtain that
\begin{align*}
    & \sigma_1^\alpha \left( \text{DICMR}_\alpha(0) - \text{DICMR}_\alpha(1) \right)\\
    \geq \quad & \left[ \widetilde{r}_\sigma^\alpha \norm{f}_{1+\alpha}^{1+\alpha} + \dfrac{\norm{f}_{1+\alpha}^{1+\alpha}}{\alpha} - \left( 1 + \dfrac{1}{\alpha} \right) \widetilde{r}_\sigma^\alpha f^\alpha(0) \right] - \dfrac{(n+p) C^f_{2\alpha}}{2np C^f_{\alpha}} + \mathcal{O}\left( (np)^{-(1+\alpha)/2} \right).
\end{align*}
\noindent Now, because of the assumption that $\widetilde{r}_\sigma \xrightarrow{P} 0$, it follows that the first term of the above lower bound goes to $\norm{f}_{1+\alpha}^{1+\alpha}/\alpha$. The fact that both the second term (the penalty function) and the third term (the remainder part) go to $0$ as both $n$ and $p$ tend to infinity is obvious. Therefore, for sufficiently large $n$ and $p$, with probability tending to $1$, we must have
\begin{equation*}
    \sigma_1^\alpha \left( \text{DICMR}_\alpha(0) - \text{DICMR}_\alpha(1) \right) \geq \norm{f}_{1+\alpha}^{1+\alpha}/\alpha > 0,
\end{equation*}
\noindent which completes the proof.

\subsection{Proof of Theorem~\ref{thm:rank-selection-consistency-zero}}\label{appendix:rank-proof-zero}

We begin with a decomposition of the difference between DICMR evaluated at rank zero and rank one as shown in~\eqref{eqn:rank-dicmr-difference}. We follow the same notations as in Section~\ref{appendix:rank-proof-one}.

Now, due to the Law of Large Numbers, it follows that as $n$ and $p$ both tend to infinity, we have
\begin{equation*}
    \dfrac{1}{np}\sum_{i=1}^n \sum_{j=1}^p (\widehat{\sigma}_\alpha^{(0)})^{-\alpha} f^\alpha\left( \dfrac{\vert X_{ij} \vert}{\widehat{\sigma}_\alpha^{(0)}} \right) = (\widehat{\sigma}_\alpha^{(0)})^{-\alpha} \norm{f}_{1+\alpha}^{1+\alpha} + \Ocal(1/\sqrt{np}).
\end{equation*}
\noindent Because of Assumptions (B1)-(B6) and an application of ``rSVDdpd'' consistency theorem as in~\cite{roy2024robustsvd} implies that $\left\vert (\widehat{\sigma}_\alpha^{(0)})^2 - \sigma_0^2 \right\vert \xrightarrow{P} 0$. Similarly, it also follows from a generalized version of Khinchin's law of large numbers that as both $n$ and $p$ tend to infinity, we have
\begin{align*}
    & \dfrac{1}{np}\sum_{i=1}^n \sum_{j=1}^p (\widehat{\sigma}_\alpha^{(1)})^{-\alpha} f^\alpha\left( \dfrac{\vert X_{ij} - \widehat{\lambda}_{1}\widehat{u}_{1i}\widehat{v}_{1j} \vert}{\widehat{\sigma}_\alpha^{(1)}} \right)\\
    = \quad & \dfrac{1}{np}\sum_{i=1}^n \sum_{j=1}^p \E\left[ \int (\widehat{\sigma}_\alpha^{(1)})^{-\alpha}\sigma_0^{-1} f^\alpha\left( \dfrac{\vert x -  \widehat{\lambda}_{1}\widehat{u}_{1i}\widehat{v}_{1j} \vert}{\widehat{\sigma}_\alpha^{(1)}} \right) f\left( \dfrac{\vert x \vert}{ \sigma_0 } \right)dx\right] + \Ocal(1/\sqrt{np})\\
    = \quad & \E\left[ \int (\widehat{\sigma}_\alpha^{(1)})^{-\alpha}\sigma_0^{-1} f^\alpha\left( \dfrac{\vert x -  \widehat{\lambda}_{1}\widehat{u}_{11}\widehat{v}_{11} \vert}{\widehat{\sigma}_\alpha^{(1)}} \right) f\left( \dfrac{\vert x \vert}{ \sigma_0 } \right)dx\right] + \Ocal(1/\sqrt{np})\\
    & \quad \text{since, the rows and the columns of } \bb{X} \text{ matrix are exchangeable}\\
    = \quad & \E\left[ (\widehat{\sigma}_\alpha^{(1)})^{-\alpha} \int f^\alpha\left( \left\vert r_\sigma z - \dfrac{\widehat{\lambda}_1\widehat{u}_{11}\widehat{v}_{11}}{\widehat{\sigma}_\alpha^{(1)}} \right\vert \right) f(\vert z\vert)dz \right] + \Ocal(1/\sqrt{np}),
\end{align*}
\noindent where $r_\sigma := \dfrac{\sigma_0}{\widehat{\sigma}_\alpha^{(1)}}$. Now note that the entries $X_{ij}$ of the data matrix can be represented as $X_{ij} = 0 . u_{1i} . v_{1j} + e_{ij}$, where $u_{1i}, v_{1j}$ are any randomly chosen but fixed vector in respective Stiefel manifold and $e_{ij}$s are independent and normally distributed errors with the same distribution as $X_{ij}$. Due to this decomposition, from the result on the consistency of the ``rSVDdpd'' estimator as given in~\cite{roy2024robustsvd}, we have $\widehat{\lambda}_1 \xrightarrow{P} 0$, as both $n$ and $p$ tend to infinity. More importantly, it also shows that the error estimate $\widehat{\sigma}_\alpha^{(1)}$ is also a consistent estimator of the original noise variance $\sigma_0$, as the first singular value and vectors asymptotically have no contribution. As a result, $r_\sigma \xrightarrow{P} 1$ as both $n$ and $p$ tend to infinity. Now, since the model density $f$ is bounded by $f(0)$ because of its decreasing nature, by an application of the Dominated Convergence Theorem (DCT), it follows that
\begin{align*}
    & \int f^\alpha\left( \left\vert r_\sigma z - \dfrac{\widehat{\lambda}_1\widehat{u}_{11}\widehat{v}_{11}}{\widehat{\sigma}_\alpha^{(1)}} \right\vert \right) f(\vert z\vert)dz\\
    = {} & \int f^\alpha\left( \left\vert z - \dfrac{\widehat{\lambda}_1\widehat{u}_{11}\widehat{v}_{11}}{\widehat{\sigma}_\alpha^{(1)}} \right\vert \right) f(\vert z\vert)dz + \Ocal(1/\sqrt{np})\\
    \leq {} & \max\left\{ \int f^\alpha\left( \left\vert z - \dfrac{\widehat{\lambda}_1\widehat{u}_{11}\widehat{v}_{11}}{\widehat{\sigma}_\alpha^{(1)}} \right\vert \right) f\left(\left\vert z - \dfrac{\widehat{\lambda}_1\widehat{u}_{11}\widehat{v}_{11}}{\widehat{\sigma}_\alpha^{(1)}} \right\vert \right)dz, \int f^\alpha\left( \left\vert z \right\vert \right) f\left(\left\vert z \right\vert \right)dz \right\} + \Ocal(1/\sqrt{np})\\
    & \qquad \text{since, } f \text{ is decreasing in the magnitude of its arguments}\\
    \leq {} & \int f^\alpha\left( \left\vert z \right\vert \right) f\left(\left\vert z \right\vert \right)dz + \Ocal(1/\sqrt{np})\\
    = {} & \norm{f}_{1+\alpha}^{1+\alpha} + \Ocal(1/\sqrt{np}).
\end{align*}
\noindent Putting everything back in Eq.~\eqref{eqn:rank-dicmr-difference} now yields that
\begin{align*}
    & \sigma_0^\alpha \left( \text{DICMR}_\alpha(0) - \text{DICMR}_\alpha(1) \right)\\
    ={} & \norm{f}_{1+\alpha}^{1+\alpha} \left[ -\dfrac{1}{\alpha} - r_\sigma^\alpha + \left( 1 + \dfrac{1}{\alpha}\right) r_\sigma^\alpha \right] - \dfrac{(n+p)}{2np}r_\sigma^\alpha \dfrac{C^f_{2\alpha}}{C^f_\alpha} + \Ocal\left( (np)^{-(1+\alpha)/2} \right).
\end{align*}
\noindent This can further be rearranged as
\begin{equation*}
    \left( r_\sigma^\alpha - 1 \right) \left[ \dfrac{\norm{f}_{1+\alpha}^{1+\alpha}}{\alpha} - \dfrac{(n+p)}{2np}\dfrac{C_{2\alpha}^f}{C_\alpha^f} \right] - \dfrac{(n+p)}{2np}\dfrac{C_{2\alpha}^f}{C_\alpha^f} + \Ocal\left( (np)^{-(1+\alpha)/2} \right).
\end{equation*}
\noindent Since $r_\sigma \xrightarrow{P} 1$, it remains bounded away from $0$ for sufficiently large $n$ and $p$. Therefore, the first term in the above sum converges in probability to $0$. The second term also converges to $0$, but from the negative side, at the rate of $(np)^{-1/2}$, as $\lim_{n,p \rightarrow \infty}(n/p) = c \in (0, \infty)$. Finally, the remainder term converges to $0$ at a faster rate $(np)^{-(1+\alpha)/2}$ for all $\alpha > 0$. Therefore, the entire difference becomes negative for all sufficiently large $n$ and $p$ if and only if the rate at which the first term goes to $0$ is smaller than the rate of the second term going to $0$.

To obtain the rate of convergence, we apply the linear regression analysis present in~\cite{ghosh2013robust} to obtain the asymptotic distribution of $r_\sigma^\alpha$. By direct calculation, one may obtain that
\begin{align*}
    J_\alpha\left( \widehat{\sigma}_1^2 \right) & = \dfrac{1}{4}\sigma_0^{-(\alpha+4)} B_\alpha^f,\\
    K_\alpha\left( \widehat{\sigma}_1^2 \right) & = \dfrac{1}{4}\sigma_0^{-(2\alpha+4)}B_{2\alpha}^f - \dfrac{1}{4}\sigma_0^{-2(2+\alpha)} (A_\alpha^f)^2,
\end{align*}
\noindent where we use the fact that as the low-rank component $\bb{L} = \bb{0}$, the expectation $\E((\sigma_\alpha^{(1)})^2) \rightarrow \sigma_0^2$ as $n, p \to \infty$. The quantities $A_\alpha^f$ and $B_\alpha^f$ are as given in~\eqref{eqn:rank-A-alpha-func}-\eqref{eqn:rank-B-alpha-func}. As a result, conditional on the estimates of the singular values and vectors, the estimated noise variance $(\widehat{\sigma}_\alpha^{(1)})^2$ has an asymptotic normal distribution as 
\begin{equation*}
    \sqrt{np}\left( (\widehat{\sigma}_\alpha^{(1)})^2 - \sigma_0^2 \right) \xrightarrow{d} \normdist\left(0, 4\sigma_0^4 \dfrac{B_{2\alpha}^f - (A_\alpha^f)^2 }{(B_\alpha^f)^2} \right).
\end{equation*} 
\noindent An application of delta method now yields the asymptotic distribution of $r_\sigma^\alpha$ as 
\begin{equation*}
    \sqrt{np}(r_\sigma^\alpha - 1) \xrightarrow{d} \normdist\left(0, \alpha^2 \dfrac{B_{2\alpha}^f - (A_\alpha^f)^2 }{(B_\alpha^f)^2} \right),
\end{equation*}
\noindent as both $n, p \to \infty$. Now fix any $t > 0$. By an application of Berry-Esseen inequality, we obtain that with probability at most $1/2t^2$,
\begin{equation*}
    (r_\sigma^\alpha - 1) > \dfrac{\alpha t}{\sqrt{np}} \dfrac{\sqrt{B_{2\alpha}^f - (A_\alpha^f)^2} }{B_\alpha^f}.
\end{equation*}
\noindent Also, if $(r_\sigma^\alpha - 1) \leq 0$, we have $\text{DICMR}_\alpha(0) < \text{DICMR}_\alpha(1)$, trivially. Therefore, for any $\alpha > 0$ and for sufficiently large $n$ and $p$, we have 
\begin{equation*}
    \text{DICMR}_\alpha(0) - \text{DICMR}_\alpha(1) < 0,
\end{equation*}
\noindent with probability at least $1/2 + \max\left\{ 0, 1/2 - 1/2t^2 \right\}$ for any $t$ such that
\begin{align*}
    \dfrac{\norm{f}_{1+\alpha}^{1+\alpha}}{\alpha} \dfrac{\alpha t}{\sqrt{np}} \dfrac{\sqrt{B_{2\alpha}^f - (A_\alpha^f)^2} }{B_\alpha^f}
    < \dfrac{(n+p)}{2np}\dfrac{C_{2\alpha}^f}{C_\alpha^f}
    \implies t < \dfrac{(1+c)}{2\sqrt{c}} \dfrac{C_{2\alpha}^f}{C_\alpha^f \norm{f}_{1+\alpha}^{1+\alpha} } \dfrac{B_\alpha^f}{\sqrt{B_{2\alpha}^f - (A_\alpha^f)^2} }.
\end{align*}
\noindent Choosing $t = (1-\epsilon)t_\alpha$ as in~\eqref{eqn:rank-t-func} implies that the DICMR is minimized for rank zero with the given probability.

\subsection{Proof of Theorem~\ref{thm:dicmr-B-robust}}

Let us denote the densities of $G_{ij}, K_{ij}$ and $G_{\epsilon, ij}$ as $g_{ij}, k_{ij}$ and $g_{\epsilon,ij} := (1-\epsilon)g_{ij} + \epsilon k_{ij}$ respectively. Then, we can express the DICMR criterion as
\begin{align}
    T(G) & := \sigma^{-\alpha} \left[ \int f^{1+\alpha} - \left( 1 + \frac{1}{\alpha} \right)\frac{1}{np}\sum_{i,j} \int f^\alpha\left(  \frac{x - \sum_{k=1}^r \lambda_k u_{ki} v_{kj} }{\sigma} \right)g_{ij}(x)dx + \frac{r(n+p)}{2np} \frac{C_{2\alpha}^f}{C_\alpha^f} \right]\nonumber\\
    & = T_1(G) \left[ \int f^{1+\alpha} - \frac{r(n+p)}{2np} \frac{C_{2\alpha}^f}{C_\alpha^f} \right] - \left(1 + \frac{1}{\alpha} \right) \frac{1}{np} \sum_{i=1}^n \sum_{j=1}^p T_{2,ij}(G)\label{eqn:rank-T-into-T1T2}
\end{align}
\noindent where $T_1(G) = \sigma^{-\alpha}$ and $T_2(G)$ is the second term of the summand. Note that, here all the parameters $\lambda_k, u_{ki}, v_{kj}$s and $\sigma$ are functionals and hence depends on $G$; but for ease of notation, we suppress this explicit dependence.

Now, consider the first term and let us denote $\partial_\epsilon$ as the von-Mises differentiation operator for functionals. It is easy to see that,
\begin{equation*}
    \partial_\epsilon T_1 = -\alpha \sigma^{-(\alpha+1)} (\partial_\epsilon \sigma).
\end{equation*}
\noindent Taking $\epsilon = 0$ and applying the boundedness of the influence function of the rSVDdpd estimator for $\sigma$, we get that $\vert \partial_\epsilon T_1\vert_{\epsilon = 0}\vert < \infty$.

Let us denote $\mu = \sum_{k=1}^r \lambda_k u_{ki}v_{kj}$, which is again a functional depending on $G$, the data distribution. With some calculation, the second term now reduces to
\begin{align}
    \vert \partial_\epsilon T_{2,ij}\vert_{\epsilon = 0}\vert 
    & = \left\vert \sigma^{-\alpha} \int f^\alpha\left( \frac{x - \mu}{\sigma} \right) (k_{ij}(x) - g_{ij}(x))dx + \partial_\epsilon\left.\left[ \sigma^{-\alpha} \int f^\alpha\left(\frac{x - \mu}{\sigma}\right) g_{ij}(x)dx \right]\right\vert_{\epsilon = 0}\right\vert \nonumber \\
    & \leq \sigma^{-\alpha} f^\alpha(0) \int (k_{ij} + g_{ij}) + \left\vert -\alpha\sigma^{-(\alpha+1)}(\partial_\epsilon \sigma)\vert_{\epsilon = 0} \int f^\alpha\left(\frac{x - \mu}{\sigma}\right) g_{ij}(x)dx \right\vert \nonumber \\
    & \qquad \qquad + \left\vert \sigma^{-(\alpha + 2)} \int \left( \frac{x - \mu}{\sigma}\right) \psi\left( \frac{x - \mu}{\sigma} \right) (-\partial_\epsilon \mu\vert_{\epsilon = 0} \sigma - (x - \mu) \partial_\epsilon \sigma\vert_{\epsilon = 0}) g_{ij}(x) \right\vert dx \label{eqn:rank-influence-proof-1}\\
    & \leq 2\sigma^{-\alpha} f^\alpha(0) + \alpha \sigma^{-(\alpha+1)} M (\partial_\epsilon \sigma)\vert_{\epsilon = 0} + \left\vert \sigma^{-(\alpha+1)} (\partial_\epsilon \sigma)\vert_{\epsilon = 0} \int \left( \frac{x - \mu}{\sigma}\right)^2 \psi\left(\frac{x - \mu}{\sigma} \right)g_{ij}(x)dx \right\vert \nonumber\\
    & \qquad \qquad + \left\vert \sigma^{-(\alpha+1)} (\partial_\epsilon \mu)\vert_{\epsilon = 0} \int \left\vert \frac{x - \mu}{\sigma} \right\vert \psi\left(\frac{x - \mu}{\sigma} \right)g_{ij}(x)dx \right\vert \label{eqn:rank-influence-proof-2}\\
    & = \leq 2\sigma^{-\alpha} f^\alpha(0) + \alpha \sigma^{-(\alpha+1)} M (\partial_\epsilon \sigma)\vert_{\epsilon = 0} + \sigma^{-(\alpha+1)} (\partial_\epsilon \sigma)\vert_{\epsilon = 0} K \nonumber \\
    & \qquad \qquad + \left\vert \sigma^{-(\alpha+1)} (\partial_\epsilon \mu)\vert_{\epsilon = 0} \int \max\left\{ 1, \left( \frac{x - \mu}{\sigma} \right)^2 \right\} \psi\left(\frac{x - \mu}{\sigma} \right)g_{ij}(x)dx \right\vert \label{eqn:rank-influence-proof-3}\\
    & \leq 2\sigma^{-\alpha} f^\alpha(0) + \alpha \sigma^{-(\alpha+1)} M (\partial_\epsilon \sigma)\vert_{\epsilon = 0} + \sigma^{-(\alpha+1)} ((\partial_\epsilon \sigma)\vert_{\epsilon = 0} + (\partial_\epsilon \mu)\vert_{\epsilon = 0}) \max\{ M, K \}\nonumber\\
    & < \infty.
\end{align}
\noindent Here, in step~\eqref{eqn:rank-influence-proof-1}, we use triangle inequality along with Assumption (A2) of~\cite{roy2024robustsvd}, i.e., the symmetry of the model density $f$. In step~\eqref{eqn:rank-influence-proof-2}, we use Assumption~\ref{assum:integral-finite} to bound $\int f^{\alpha}(\cdot)g_{ij}(\cdot)dx$ by a quantity $M$, which may depend on $\mu$ and $\sigma$. In step~\eqref{eqn:rank-influence-proof-3}, we make use of Assumption (A3) of~\cite{roy2024robustsvd} to bound $x^2\psi(x)$ and also use the fact that $g_{ij}$ is a density that integrates to $1$. Finally at the last step, we use the fact that when $\alpha > 0$, the rSVDdpd estimator has bounded influence function.

Combining the boundedness of the derivatives of $T_1(G)$ and $T_{2,ij}(G)$, and using expression~\eqref{eqn:rank-T-into-T1T2}, the result now follows.

\pagebreak
\section{Appendix: Detailed Simulation Results}\label{appendix:rank-sim-detail}

In this section, we present the detailed tables containing the simulation results for robust rank estimation using existing model selection criteria and the proposed DICMR. For the case of penalized methods, the robust variants are indicated by the word ``rsvd'' in brackets in the first columns of Tables~\ref{tab:rank-result-setA-svd1}-\ref{tab:rank-result-setD-svd2}. For the cross-validation techniques, the words ``MSE'' (mean squared error), ``MAE'' (mean absolute error), ``MAD'' (median absolute deviation about median) in parentheses are used to indicate the metric uses to aggregate the cross-validation errors from different holdout sets.

\begin{table}[htbp]
    \centering
    \caption{Proportion of exact (overestimation in brackets) for different rank-estimation methods when using classical SVD for the simulation setting of $50 \times 40$-dimensional matrix with equal singular values.}
    \label{tab:rank-result-setA-svd1}
    \resizebox*{\textwidth}{!}{
\begin{tabular}{@{}lrrrrrrrrrrrr@{}}
        \toprule
        \textbf{Method} & S01 & S02 & S03 & S11 & S12 & S13 & S21 & S22 & S23 & S31 & S32 & S33\\
        \midrule
        Elbow & $0.62$ & $0.62$ & $0.36$ & $0.02$ & $0.03$ & $0.02$ & $0.05$ & $0.03$ & $0.04$ & $0.01$ & $0.02$ & $0.01$ \\
         & $(0.00)$ & $(0.00)$ & $(0.05)$ & $(0.15)$ & $(0.10)$ & $(0.15)$ & $(0.11)$ & $(0.17)$ & $(0.14)$ & $(0.12)$ & $(0.09)$ & $(0.11)$\\
        AIC & $0.09$ & $0.00$ & $0.00$ & $0.10$ & $0.04$ & $0.04$ & $0.06$ & $0.05$ & $0.05$ & $0.13$ & $0.12$ & $0.15$ \\
         & $(0.00)$ & $(0.00)$ & $(0.00)$ & $(0.11)$ & $(0.06)$ & $(0.03)$ & $(0.07)$ & $(0.07)$ & $(0.07)$ & $(0.49)$ & $(0.50)$ & $(0.47)$\\
        BIC & $0.00$ & $0.00$ & $0.00$ & $0.00$ & $0.00$ & $0.00$ & $0.00$ & $0.00$ & $0.00$ & $0.00$ & $0.00$ & $0.00$ \\
         & $(0.00)$ & $(0.00)$ & $(0.00)$ & $(0.00)$ & $(0.00)$ & $(0.00)$ & $(0.00)$ & $(0.00)$ & $(0.00)$ & $(0.00)$ & $(0.00)$ & $(0.00)$\\
        PC1 & $0.00$ & $0.00$ & $0.00$ & $0.00$ & $0.00$ & $0.00$ & $0.00$ & $0.00$ & $0.00$ & $0.00$ & $0.00$ & $0.00$ \\
\bottomrule
    \end{tabular}}
\end{table}

\begin{table}[htbp]
    \centering
    \caption{Bias (RMSE in brackets) for different rank-estimation methods when using classical SVD for the simulation setting of $50 \times 40$-dimensional matrix with equal singular values.}
    \label{tab:rank-result-setA-svd2}
    \resizebox*{\textwidth}{!}{
    \begin{tabular}{@{}lrrrrrrrrrrrr@{}}
        \toprule
        \textbf{Method} & S01 & S02 & S03 & S11 & S12 & S13 & S21 & S22 & S23 & S31 & S32 & S33\\
        \midrule
        Elbow & $-0.38$ & $-0.38$ & $-1.07$ & $-4.74$ & $-5.38$ & $-4.92$ & $-5.24$ & $-4.87$ & $-5.12$ & $-5.28$ & $-5.70$ & $-5.56$ \\
         & $(0.62)$ & $(0.62)$ & $(2.21)$ & $(6.28)$ & $(6.65)$ & $(6.57)$ & $(6.94)$ & $(7.14)$ & $(7.24)$ & $(6.70)$ & $(7.04)$ & $(7.01)$\\
        AIC & $-8.03$ & $-9.00$ & $-9.00$ & $-2.90$ & $-4.08$ & $-4.79$ & $-3.48$ & $-3.53$ & $-3.55$ & $ 0.58$ & $ 0.54$ & $ 0.51$ \\
         & $(8.49)$ & $(9.00)$ & $(9.00)$ & $(4.24)$ & $(4.95)$ & $(5.44)$ & $(4.25)$ & $(4.30)$ & $(4.37)$ & $(2.83)$ & $(2.81)$ & $(2.79)$\\
        BIC & $-9.00$ & $-9.00$ & $-9.00$ & $-9.00$ & $-9.00$ & $-9.00$ & $-9.00$ & $-9.00$ & $-9.00$ & $-9.00$ & $-9.00$ & $-9.00$ \\
         & $(9.00)$ & $(9.00)$ & $(9.00)$ & $(9.00)$ & $(9.00)$ & $(9.00)$ & $(9.00)$ & $(9.00)$ & $(9.00)$ & $(9.00)$ & $(9.00)$ & $(9.00)$\\
        PC1 & $-9.00$ & $-9.00$ & $-9.00$ & $-8.37$ & $-8.63$ & $-8.79$ & $-8.74$ & $-8.75$ & $-8.77$ & $-8.10$ & $-8.08$ & $-8.11$ \\
\bottomrule
    \end{tabular}}
\end{table}

\begin{table}[htbp]
    \centering
    \caption{Proportion of exact (overestimation in brackets) for different rank-estimation methods when using classical SVD for the simulation setting of $50 \times 40$-dimensional matrix with decreasing singular values.}
    \label{tab:rank-result-setB-svd1}
    \resizebox*{\textwidth}{!}{
    \begin{tabular}{@{}lrrrrrrrrrrrr@{}}
        \toprule
        \textbf{Method} & S01 & S02 & S03 & S11 & S12 & S13 & S21 & S22 & S23 & S31 & S32 & S33\\
        \midrule
        Elbow & $1.00$ & $0.65$ & $0.13$ & $0.02$ & $0.03$ & $0.01$ & $0.05$ & $0.05$ & $0.04$ & $0.03$ & $0.03$ & $0.02$ \\
         & $(0.00)$ & $(0.01)$ & $(0.02)$ & $(0.16)$ & $(0.14)$ & $(0.17)$ & $(0.15)$ & $(0.18)$ & $(0.17)$ & $(0.14)$ & $(0.13)$ & $(0.13)$\\
        AIC & $0.00$ & $0.00$ & $0.00$ & $0.03$ & $0.04$ & $0.15$ & $0.03$ & $0.02$ & $0.01$ & $0.00$ & $0.00$ & $0.00$ \\
         & $(0.00)$ & $(0.00)$ & $(0.00)$ & $(0.95)$ & $(0.89)$ & $(0.72)$ & $(0.96)$ & $(0.97)$ & $(0.98)$ & $(1.00)$ & $(1.00)$ & $(1.00)$\\
        BIC & $0.00$ & $0.00$ & $0.00$ & $0.00$ & $0.00$ & $0.00$ & $0.00$ & $0.00$ & $0.00$ & $0.00$ & $0.00$ & $0.00$ \\
         & $(0.00)$ & $(0.00)$ & $(0.00)$ & $(0.00)$ & $(0.00)$ & $(0.00)$ & $(0.00)$ & $(0.00)$ & $(0.00)$ & $(0.00)$ & $(0.00)$ & $(0.00)$\\
        PC1 & $0.00$ & $0.00$ & $0.00$ & $0.00$ & $0.02$ & $0.02$ & $0.02$ & $0.03$ & $0.03$ & $0.06$ & $0.06$ & $0.07$ \\
\bottomrule
    \end{tabular}}
\end{table}

\begin{table}[htbp]
    \centering
    \caption{Bias (RMSE in brackets) for different rank-estimation methods when using classical SVD for the simulation setting of $50 \times 40$-dimensional matrix with decreasing singular values.}
    \label{tab:rank-result-setB-svd2}
    \resizebox*{\textwidth}{!}{
    \begin{tabular}{@{}lrrrrrrrrrrrr@{}}
        \toprule
        \textbf{Method} & S01 & S02 & S03 & S11 & S12 & S13 & S21 & S22 & S23 & S31 & S32 & S33\\
        \midrule
        Elbow & $ 0.00$ & $-1.39$ & $-4.87$ & $-4.82$ & $-5.06$ & $-5.05$ & $-4.74$ & $-4.61$ & $-4.52$ & $-4.95$ & $-4.85$ & $-5.13$ \\
         & $(0.00)$ & $(2.93)$ & $(5.89)$ & $(6.43)$ & $(6.49)$ & $(6.89)$ & $(6.69)$ & $(6.98)$ & $(6.92)$ & $(6.61)$ & $(6.59)$ & $(6.74)$\\
        AIC & $-2.48$ & $-9.00$ & $-9.00$ & $ 4.44$ & $ 3.30$ & $ 2.55$ & $ 5.07$ & $ 5.12$ & $ 5.01$ & $ 8.99$ & $ 8.98$ & $ 9.04$ \\
         & $(2.53)$ & $(9.00)$ & $(9.00)$ & $(5.29)$ & $(4.42)$ & $(3.96)$ & $(5.71)$ & $(5.74)$ & $(5.63)$ & $(9.08)$ & $(9.08)$ & $(9.13)$\\
        BIC & $-9.00$ & $-9.00$ & $-9.00$ & $-8.90$ & $-8.94$ & $-8.99$ & $-9.00$ & $-9.00$ & $-9.00$ & $-8.99$ & $-8.99$ & $-8.99$ \\
         & $(9.00)$ & $(9.00)$ & $(9.00)$ & $(8.92)$ & $(8.95)$ & $(8.99)$ & $(9.00)$ & $(9.00)$ & $(9.00)$ & $(8.99)$ & $(8.99)$ & $(8.99)$\\
        PC1 & $-6.60$ & $-9.00$ & $-9.00$ & $-4.76$ & $-5.52$ & $-6.07$ & $-5.27$ & $-5.41$ & $-5.42$ & $-3.11$ & $-3.09$ & $-3.05$ \\
\bottomrule
    \end{tabular}}
\end{table}

\begin{table}[htbp]
    \centering
    \caption{Proportion of exact (overestimation in brackets) for different rank-estimation methods when using robust SVD for the simulation setting of $50 \times 40$-dimensional matrix with equal singular values.}
    \label{tab:rank-result-setC-svd1}
    \resizebox*{\textwidth}{!}{
    \begin{tabular}{@{}lrrrrrrrrrrrr@{}}
        \toprule
        \textbf{Method} & S01 & S02 & S03 & S11 & S12 & S13 & S21 & S22 & S23 & S31 & S32 & S33\\
        \midrule
        Elbow (rsvd) & $0.41$ & $0.54$ & $0.30$ & $0.44$ & $0.61$ & $0.26$ & $0.42$ & $0.55$ & $0.12$ & $0.69$ & $0.50$ & $0.04$ \\
         & $(0.00)$ & $(0.00)$ & $(0.04)$ & $(0.00)$ & $(0.00)$ & $(0.01)$ & $(0.00)$ & $(0.00)$ & $(0.05)$ & $(0.00)$ & $(0.00)$ & $(0.00)$\\
        AIC (rsvd) & $0.13$ & $0.00$ & $0.00$ & $0.00$ & $0.00$ & $0.00$ & $0.00$ & $0.00$ & $0.00$ & $0.00$ & $0.00$ & $0.00$ \\
         & $(0.00)$ & $(0.00)$ & $(0.00)$ & $(0.00)$ & $(0.00)$ & $(0.00)$ & $(0.00)$ & $(0.00)$ & $(0.00)$ & $(0.00)$ & $(0.00)$ & $(0.00)$\\
        BIC (rsvd) & $0.00$ & $0.00$ & $0.00$ & $0.00$ & $0.00$ & $0.00$ & $0.00$ & $0.00$ & $0.00$ & $0.00$ & $0.00$ & $0.00$ \\
         & $(0.00)$ & $(0.00)$ & $(0.00)$ & $(0.00)$ & $(0.00)$ & $(0.00)$ & $(0.00)$ & $(0.00)$ & $(0.00)$ & $(0.00)$ & $(0.00)$ & $(0.00)$\\
        PC1 (rsvd) & $0.00$ & $0.00$ & $0.00$ & $0.00$ & $0.00$ & $0.00$ & $0.00$ & $0.00$ & $0.00$ & $0.00$ & $0.00$ & $0.00$ \\
\bottomrule
    \end{tabular}}
\end{table}

\begin{table}[htbp]
    \centering
    \caption{Bias (RMSE in brackets) for different rank-estimation methods when using robust SVD for the simulation setting of $50 \times 40$-dimensional matrix with equal singular values.}
    \label{tab:rank-result-setC-svd2}
    \resizebox*{\textwidth}{!}{
    \begin{tabular}{@{}lrrrrrrrrrrrr@{}}
        \toprule
        \textbf{Method} & S01 & S02 & S03 & S11 & S12 & S13 & S21 & S22 & S23 & S31 & S32 & S33\\
        \midrule
        Elbow (rsvd) & $-0.59$ & $-0.54$ & $-2.11$ & $-0.56$ & $-0.55$ & $-3.64$ & $-0.66$ & $-0.70$ & $-4.16$ & $-0.55$ & $-2.10$ & $-5.90$ \\
         & $(0.77)$ & $(1.12)$ & $(3.73)$ & $(0.75)$ & $(1.41)$ & $(5.16)$ & $(1.17)$ & $(1.70)$ & $(5.64)$ & $(1.65)$ & $(3.73)$ & $(6.62)$\\
        AIC (rsvd) & $-7.40$ & $-9.00$ & $-9.00$ & $-8.76$ & $-9.00$ & $-9.00$ & $-8.68$ & $-9.00$ & $-9.00$ & $-8.74$ & $-9.00$ & $-8.99$ \\
         & $(7.97)$ & $(9.00)$ & $(9.00)$ & $(8.78)$ & $(9.00)$ & $(9.00)$ & $(8.70)$ & $(9.00)$ & $(9.00)$ & $(8.75)$ & $(9.00)$ & $(8.99)$\\
        BIC (rsvd) & $-9.00$ & $-9.00$ & $-9.00$ & $-9.00$ & $-9.00$ & $-9.00$ & $-9.00$ & $-9.00$ & $-9.00$ & $-9.00$ & $-9.00$ & $-9.00$ \\
         & $(9.00)$ & $(9.00)$ & $(9.00)$ & $(9.00)$ & $(9.00)$ & $(9.00)$ & $(9.00)$ & $(9.00)$ & $(9.00)$ & $(9.00)$ & $(9.00)$ & $(9.00)$\\
        PC1 (rsvd) & $-9.00$ & $-9.00$ & $-9.00$ & $-9.00$ & $-9.00$ & $-9.00$ & $-9.00$ & $-9.00$ & $-9.00$ & $-9.00$ & $-9.00$ & $-9.00$ \\
\bottomrule
    \end{tabular}}
\end{table}

\begin{table}[htbp]
    \centering
    \caption{Proportion of exact (overestimation in brackets) for different rank-estimation methods when using robust SVD for the simulation setting of $50 \times 40$-dimensional matrix with decreasing singular values.}
    \label{tab:rank-result-setD-svd1}
    \resizebox*{\textwidth}{!}{
    \begin{tabular}{@{}lrrrrrrrrrrrr@{}}
        \toprule
        \textbf{Method} & S01 & S02 & S03 & S11 & S12 & S13 & S21 & S22 & S23 & S31 & S32 & S33\\
        \midrule
        Elbow (rsvd) & $1.00$ & $0.55$ & $0.09$ & $0.97$ & $0.36$ & $0.03$ & $0.93$ & $0.32$ & $0.01$ & $0.86$ & $0.11$ & $0.04$ \\
         & $(0.00)$ & $(0.00)$ & $(0.02)$ & $(0.00)$ & $(0.00)$ & $(0.02)$ & $(0.00)$ & $(0.00)$ & $(0.02)$ & $(0.00)$ & $(0.01)$ & $(0.02)$\\
        AIC (rsvd) & $0.00$ & $0.00$ & $0.00$ & $0.00$ & $0.00$ & $0.00$ & $0.00$ & $0.00$ & $0.00$ & $0.00$ & $0.00$ & $0.00$ \\
         & $(0.00)$ & $(0.00)$ & $(0.00)$ & $(0.00)$ & $(0.00)$ & $(0.00)$ & $(0.00)$ & $(0.00)$ & $(0.00)$ & $(0.00)$ & $(0.00)$ & $(0.00)$\\
        BIC (rsvd) & $0.00$ & $0.00$ & $0.00$ & $0.00$ & $0.00$ & $0.00$ & $0.00$ & $0.00$ & $0.00$ & $0.00$ & $0.00$ & $0.00$ \\
         & $(0.00)$ & $(0.00)$ & $(0.00)$ & $(0.00)$ & $(0.00)$ & $(0.00)$ & $(0.00)$ & $(0.00)$ & $(0.00)$ & $(0.00)$ & $(0.00)$ & $(0.00)$\\
        PC1 (rsvd) & $0.00$ & $0.00$ & $0.00$ & $0.00$ & $0.00$ & $0.00$ & $0.00$ & $0.00$ & $0.00$ & $0.00$ & $0.00$ & $0.00$ \\
\bottomrule
    \end{tabular}}
\end{table}

\begin{table}[htbp]
    \centering
    \caption{Bias (RMSE in brackets) for different rank-estimation methods when using robust SVD for the simulation setting of $50 \times 40$-dimensional matrix with decreasing singular values.}
    \label{tab:rank-result-setD-svd2}
    \resizebox*{\textwidth}{!}{
    \begin{tabular}{@{}lrrrrrrrrrrrr@{}}
        \toprule
        \textbf{Method} & S01 & S02 & S03 & S11 & S12 & S13 & S21 & S22 & S23 & S31 & S32 & S33\\
        \midrule
        Elbow (rsvd) & $ 0.00$ & $-2.12$ & $-5.26$ & $-0.03$ & $-3.04$ & $-5.69$ & $-0.07$ & $-3.89$ & $-5.99$ & $-0.27$ & $-5.81$ & $-6.43$ \\
         & $(0.00)$ & $(3.79)$ & $(6.19)$ & $(0.17)$ & $(4.54)$ & $(6.45)$ & $(0.26)$ & $(5.24)$ & $(6.62)$ & $(1.12)$ & $(6.69)$ & $(7.05)$\\
        AIC (rsvd) & $-2.47$ & $-9.00$ & $-9.00$ & $-3.00$ & $-9.00$ & $-9.00$ & $-3.45$ & $-8.97$ & $-9.00$ & $-4.19$ & $-8.49$ & $-8.87$ \\
         & $(2.52)$ & $(9.00)$ & $(9.00)$ & $(3.03)$ & $(9.00)$ & $(9.00)$ & $(3.49)$ & $(8.97)$ & $(9.00)$ & $(4.23)$ & $(8.51)$ & $(8.88)$\\
        BIC (rsvd) & $-9.00$ & $-9.00$ & $-9.00$ & $-9.00$ & $-9.00$ & $-9.00$ & $-9.00$ & $-9.00$ & $-9.00$ & $-9.00$ & $-9.00$ & $-9.00$ \\
         & $(9.00)$ & $(9.00)$ & $(9.00)$ & $(9.00)$ & $(9.00)$ & $(9.00)$ & $(9.00)$ & $(9.00)$ & $(9.00)$ & $(9.00)$ & $(9.00)$ & $(9.00)$\\
        PC1 (rsvd) & $-6.65$ & $-9.00$ & $-9.00$ & $-6.96$ & $-9.00$ & $-9.00$ & $-7.12$ & $-9.00$ & $-9.00$ & $-7.56$ & $-9.00$ & $-9.00$ \\
\bottomrule
    \end{tabular}}
\end{table}

\end{document}